\documentclass[table, screen, nonacm, acmsmall]{acmart}

\usepackage{tabularx}
\usepackage{booktabs}
\usepackage{graphicx}
\usepackage{hyperref}
\usepackage{hhline}
\usepackage{geometry}
\usepackage{tabularray}
\usepackage{subcaption}
\usepackage{enumitem}
\usepackage{float} 
\usepackage{soul,color}
\usepackage[utf8]{inputenc}
\usepackage{xcolor}
\usepackage{pbox}
\usepackage{multirow}
\usepackage{array}
\usepackage{makecell}
\usepackage{afterpage} 
\usepackage[title]{appendix}
\usepackage{ragged2e}

\newcommand{\edit}[1]{\textcolor{black}{#1}}

\newcolumntype{L}[1]{>{\raggedright\let\newline\\\arraybackslash\hspace{0pt}}m{#1}}
\newcolumntype{C}[1]{>{\centering\let\newline\\\arraybackslash\hspace{0pt}}m{#1}}
\newcolumntype{R}[1]{>{\raggedleft\let\newline\\\arraybackslash\hspace{0pt}}m{#1}}
\AtBeginDocument{%
 }

\setcopyright{rightsretained}
\acmJournal{PACMHCI}
\acmYear{2024} 
\acmVolume{8} 
\acmNumber{CSCW2} 
\acmArticle{422} 
\acmMonth{11}
\acmDOI{10.1145/3686961}

\makeatletter
\gdef\@copyrightpermission{
  \begin{minipage}{0.2\columnwidth}
    \href{https://creativecommons.org/licenses/by-nc-nd/4.0/}{\includegraphics[width=0.90\textwidth]{Figures/4ACM-CC-by-nc-nd-88x31.eps}}
  \end{minipage}\hfill
  \begin{minipage}{0.8\columnwidth}
    \href{https://creativecommons.org/licenses/by-nc-nd/4.0/}{This work is licensed under a Creative Commons Attribution-NonCommercial-NoDerivs International 4.0 License.}
  \end{minipage}
  \vspace{5pt}
}
\makeatother

\begin{document}

\title[Adolescent Social Media Use]{Teen Talk: The Good, the Bad, and the Neutral of Adolescent Social Media Use}

\author{Abdulmalik Alluhidan}
\email{abdulmalik.s.alluhidan@vanderbilt.edu}
\orcid{0000-0001-8948-3923}
\affiliation{
  \institution{Vanderbilt University}
  \city{Nashville}
  \state{Tennessee}
  \country{USA}
}

\author{Mamtaj Akter}
\email{mamtaj.akter@nyit.edu}
\orcid{0000-0002-5692-9252}
\affiliation{
  \institution{New York Institute of Technology}
  \city{New York}
  \state{New York}
  \country{USA}
}

\author{Ashwaq Alsoubai}
\email{atalsoubai@kau.edu.sa}
\orcid{0000-0003-1569-9662}
\affiliation{
  \institution{King AbdulAziz University}
  \city{Jeddah}
  \country{KSA}
}

\author{Jinkyung Katie Park}
\email{jinkyup@clemson.edu}
\orcid{0000-0002-0804-832X}
\affiliation{
  \institution{Clemson University}
  \city{Clemson}
  \state{South Carolina}
  \country{USA}
}

\author{Pamela Wisniewski}
\email{pamela.wisniewski@vanderbilt.edu}
\orcid{0000-0002-6223-1029}
\affiliation{
  \institution{Vanderbilt University}
  \city{Nashville}
  \state{Tennessee}
  \country{USA}
}
\received{July 2023}
\received[revised]{January 2024}
\received[accepted]{March 2024}
\renewcommand{\shortauthors}{Abdulmalik Alluhidan et al.}

\begin{abstract}
The debate on whether social media has a net positive or negative effect on youth is ongoing. Therefore, we conducted a thematic analysis on \edit{2,061 posts made by 1,038 adolescents} aged 15-17 on an online peer-support platform to investigate the ways in which these teens discussed popular social media platforms in their posts and to identify differences in their experiences across platforms. Our findings revealed four main emergent themes for the ways in which social media was discussed: 1) Sharing negative experiences or outcomes of social media use \edit{(58\%, \textit{n} = 1,095),}  2) Attempts to connect with others \edit{(45\%, \textit{n} = 922)}, 3) Highlighting the positive side of social media use \edit{(20\%, \textit{n} = 409)}, and 4) Seeking information \edit{(20\%, \textit{n} = 491)}. \edit{Overall, while sharing about negative experiences was more prominent, teens also discussed balanced perspectives of connection-seeking, positive experiences, and information support on social media that should not be discounted.} Moreover, we found statistical significance for how these experiences differed across social media platforms. For instance, teens were most likely to seek romantic relationships on Snapchat and self-promote on YouTube. Meanwhile, Instagram was mentioned most frequently for body shaming, and Facebook was the most commonly discussed platform for privacy violations (mostly from parents). The key takeaway from our study is that the benefits and drawbacks of teens' social media usage can co-exist and net effects (positive or negative) can vary across different teens across various contexts. As such, we advocate for mitigating the negative experiences and outcomes of social media use as voiced by teens, to improve, rather than limit or restrict, their overall social media experience. We do this by taking an affordance perspective that aims to promote the digital well-being and online safety of youth ``by design.''
\end{abstract}



\maketitle

\section{Introduction}


Teens today are unique in that they have grown up in a society that is becoming more digitized and heavily relies on screens \cite{orben2020teenagers}. Having access to smartphones played an important role in this trend as the percentage of teens who have access to smartphones increased from 73\% in 2015 to reach 95\% in 2022 \cite{vogels2022teens}. By 2022, nearly half of U.S. teens were online almost constantly; more than 75\% of teens are on social media; about one-in-five teens use social media almost constantly; and more than half of teens said it would be hard to give up social media ~\cite{vogels2022teens}. As such, social media usage has become a fundamental and inseparable aspect of the lives of teenagers. In addition, the social media platforms teens use are becoming more varied; out of seven platform options that were presented to teens to choose from, 71\% of them said they use more than one social media platform \cite{lenhart2015teens}. Among them, YouTube, TikTok, Instagram, and Snapchat were the most popular, with Facebook usage dropping significantly compared to previous years ~\cite{vogels2022teens}. Teens often use different social media platforms for various purposes due to a combination of factors, including social factors such as joining popular platforms to interact with their friends, who were already members of the platform~\cite{ahn2011effect}, psychological factors such as fear of missing out on trends or events~\cite{kamaruddin2022relationship}, and technological factors including unique features and tools that are catered for teens' needs such as short videos on TikTok or temporary content on Snapchat~\cite{wulandari2021utilization,cristofaro2023initial}. Therefore, understanding teens' different experiences across these social media platforms becomes of great importance for different stakeholders including parents, educators, and policymakers to better support teens, provide appropriate guidance, and ensure their positive and safe engagement in the digital world.  




According to Pew Research, teens utilize social media to talk about a variety of subjects, such as their family life, successes, emotions, and romantic life \cite{anderson2022post}. Also, teens employ social media to seek support because it allows them to connect with like-minded individuals without worrying about being stigmatized or judged~\cite{sanger2022social, vogels_teens_2023}. Social media enables teens to establish networks and communities that offer support for discussing sensitive subjects while giving them control over how much information they disclose to others \cite{sanger2022social}. This is particularly true on pseudo-anonymous peer support platforms, as they offer many advantages, such as emotional support, access to advice, meaningful social interactions, and the opportunity for individuals to share and articulate their emotions and perspectives openly with strangers~\cite{rayland2023social}. At the same time, many researchers (e.g., ~\cite{turkle2023always,ayinmoro_sexting_2020, brinkley_sending_2017, chaudhary_sexting_2018, temple_brief_2014}) and the news media~\cite{WSJ_2022, USA_2023} 
have surfaced serious concerns regarding the negative impact social media may have on the mental health, well-being, and online safety of teens. For instance, concerns over the negative effects of social media use on adolescents spurred after The Wall Street Journal revealed internal reports from Facebook suggesting that Facebook was well aware of the adverse impacts of their platform on teens such as pushing self-harm posts and spreading misinformation owing to algorithms embedded on the platform~\cite{WSJ_2022}.  
As a result, several mass civil action lawsuits ~\cite{cbsnews_more_2023, wkrn_tn_2023} and nationwide legal reform efforts (e.g., Kids Online Safety Act~\cite{sen_blumenthal_text_2022}) have come to the forefront as a way to proactively shield youth from the detrimental effects of social media. At the same time, other members of the research community have pushed back to suggest that the fear-based narratives and `moral panic' ~\cite{naezer_risky_2018, radesky_moral_2022} around social media's impact on youth may be overblown \cite{choi_adolescent_2019}. For instance, Odgers et al. ~\cite{odgers2022engaging} poignantly urged us to move beyond arguing over the positive versus negative effects of social media use on youth; rather, it is our job to maximize the benefits and mitigate risks. 


Therefore, as we stand at the cusp of this critical and ongoing debate, it is imperative that we present compelling and \textit{balanced} evidence regarding the effects of social media on our future generations in order to inform future policies and practices that will shape how youth engage with social media for years to come. It is even more important that we do so based on the lived experiences and voices of the youth themselves. In this paper, we accomplish this goal by conducting a qualitative thematic analysis of \edit{2,061} posts made by \edit{1,038} teens (ages 15-17) on an online peer support platform, where they mentioned one or more of the top ten most popular social media platforms used in the U.S. by teens (i.e., Twitter, Facebook, Snapchat, YouTube, TikTok, WhatsApp, Reddit, Twitch, and Tumblr ~\cite{vogels2022teens}). We qualitatively coded these posts and conducted statistical analyses to determine significant differences in our codes based on the frequencies in which different social media platforms were present in our qualitative themes. As such, we answered the following high-level research questions:


\begin{itemize}
  \item \textbf{RQ1:} \textit{In what ways do youth discuss popular social media platforms when seeking support?}
   \item \textbf{RQ2:} \textit{When youth discuss their experiences on different social media platforms, are there significant differences in the subthemes that emerge within their posts?}
\end{itemize}


 Overall, we found that teens most often posted regarding Instagram (35\% of posts), Snapchat (\edit{20}\%), and YouTube (\edit{14}\%), respectively. Teens primarily utilized the peer-support platform to talk about their \textit{negative experiences when engaging with social media} (\edit{43}\% of posts). These negative experiences included social drama, cyberbullying, body shaming, harmful content viewing, and interpersonal privacy violations. Based on relative percentages of overall posts, teens called out Instagram for body-shaming and cyberbullying while Facebook for interpersonal privacy violations (mostly by parents). 
 Teens also identified several negative outcomes of social media use (\edit{15}\% of posts), including harm to their self-esteem, arousal of anger, mental health issues, and problems with time management due to addictive behaviors. Instagram appeared to be most frequently and significantly associated with harm to one's self-esteem, while YouTube was attributed to being the most time addictive. The next most common theme teens discussed was how they \textit{sought connection through social media} (\edit{45}\% of posts). YouTube was leveraged most often for self-promotion, while teens sought intimate relationships significantly more frequently on Snapchat and offered or sought emotional support on Facebook and WhatsApp. Additionally, teens mentioned the \textit{positive side of engaging on social media} (\edit{20}\% of posts). For instance, they identified Snapchat, Twitter and Facebook as platforms that helped them more often cope with mental distress (often through distraction), and 
 \edit{Instagram} was the most frequently cited platform where they encountered inspiration or supportive content. Finally, teens \textit{sought informational support} (\edit{20}\%) regarding social media platforms, such as verifying the legitimacy of content posted, how to effectively use the features of certain platforms, and seeking technical support. Content-related discussions were relatively more frequent with YouTube (less so on Instagram), while how to use the platform and tech support was most common with Twitter and Instagram, not on YouTube, respectively. 

\edit{With the continuous increase in teens' use of social media, exploring the topic with a broad perspective of both positive and negative effects is essential.} Conducting a comprehensive examination of adolescent posts on an online peer support platform provided us with a unique opportunity to discover nuances of how teens discuss their lived experiences across different social media platforms. While most of the previous works examined teens' online behaviors either irrespective of social media platforms or by relying on a single platform, ours is one of the first to pinpoint some of the potential strengths and weaknesses of different social media platforms \edit{by analyzing a unique dataset through both qualitative and quantitative methods}. To some extent, our findings can be explained by the underlying affordances, the perceived features of social media platforms that enable or constrain users' behavior on the platform~\cite{ronzhyn2022defining}, 
of popular social media platforms. For instance, it makes sense that Instagram, known for sharing photos, was associated most strongly with concerns about body-shaming and poor self-esteem. Alternatively, YouTube being associated with self-promotion and time management concerns speaks to its emphasis on easy-to-use content sharing, compared to the other social media platforms focused more on social network connections. These results highlight how social media should not be treated as a monolith as different social media platforms have unique affordances that shape the overall user experience of our youth. Therefore, targeted risk mitigation strategies need to be designed to address the unique affordances and social norms of each platform. As such, this research is valuable to the CSCW community and beyond, \edit{as researchers,  designers, policymakers, and practitioners can benefit from this knowledge in developing and implementing interventions} that take into account the changing patterns of engagement across different platforms. \edit{A key strength of our work is in demonstrating for future research the importance of examining the lived experience of teens across various social media platforms and doing so based on digital trace data in places where they are actively seeking support based on their experiences.} Overall, this research makes the following novel research contributions to the online safety and digital youth literature:
\begin{itemize}
  \item Provides an understanding of teens' lived experiences on social media platforms through the analyses of \edit{a unique dataset of 2,061 pseudo-anonymous support-seeking posts shared by 1,038} teens.
  \item Pinpoints \edit{a broad perspective of both} the potential strengths and weaknesses of different social media platforms through the lens of social media affordances. \edit{It is one of the first to take an affordances perspective to better understand youths' differentiated social media experiences across different platforms.}
  \item Provides evidence-based implications for researchers and designers to \edit{proactively} develop interventions that reflect the unique affordances and social norms of each social platform. It also suggests implications for policies and practices that can shape future generations' positive engagement with social media. 
\end{itemize}

\section{Background}
In this section, we introduce the relevant literature regarding the debates over the net effects of social media use on adolescents, analyzing support-seeking posts to understand teens' lived social media experiences, and affordances perspectives to understand teens' differing social media experiences. 


\subsection{Debating the `Net Effects' of Social Media Use on Adolescents}

Prior research has documented various benefits of teens' social media usage, such as connecting and socializing to break their isolation~\cite{maheux2021grateful}, learning~\cite{asterhan2017teenage}, and building support networks~\cite{sanger2022social}.  These positive effects were found to be particularly salient during and after the physical isolation due to the global pandemic ~\cite{haddock2022positive}. For instance, Maheux et al.~\cite{maheux2021grateful} conducted a longitudinal study with 704 adolescents to examine the impact of social media use on their gratitude before and during the pandemic over 15 months. In this study, a significant positive association was found between adolescents' social media usage for meaningful conversations with their friends and their gratitude over time. 
Additionally, social media facilitated online peer support, which also benefits youth by providing users the ability to connect with similar others without fear of stigma or judgment, to create supportive networks and communities to discuss sensitive topics while choosing how much to share with others~\cite{sanger2022social}. Online peer support platforms were found to offer benefits for teens such as emotional support, the availability of advice, engagement in valuable social interactions, and the opportunity to disclose and express feelings and views~\cite{rayland2023social}. Especially, online peer support for individuals with \textit{mental health challenges} experienced increased empowerment, self-efficacy, and better management of depression, enhanced coping strategies, and reduced social isolation ~\cite{melling2011online}. Overall, it is important to note that while social media provides a wide array of benefits, responsible and mindful use is crucial to have a safer engagement online.  


Therefore, over the years, a major number of prior works have extensively studied the negative impact of social media usage on adolescents, shedding light on its multifaceted consequences~\cite{valkenburg2022social,paat2021digital}. For instance, researchers have consistently pointed to numerous concerns related to the relationship between teens' social media usage and their mental health issues, such as increased feelings of loneliness, anxiety, and depression~\cite{keles2020systematic}. Given the prevalence of cyberbullying among adolescents due to the usage of social media platforms, several studies have primarily focused on examining the detrimental impact of cyberbullying on teens, uncovering how these experiences had severe psychological effects on teens such as increased depression, anxiety, and suicide ideation~\cite{kwan2020cyberbullying, levine2022gendered, doumas2023witnessing}. 
\edit{
In addition, several studies have explored the negative impact of social media through the lens of social comparison, a process where individuals assess their opinions, values, achievements, and abilities in relation to those of others \cite{powdthavee2014social}. For example, past research found that using Instagram may heighten social comparison, subsequently leading to a decrease in self-esteem \cite{jiang2020effects}. Further studies have also indicated that the use of Instagram could lead to feelings of insufficiency due to upward social comparison \cite{hwnag2019social}. Other researchers have looked at the negative aspects of social comparison shedding light on other platforms. For example, Vogel et al \cite{vogel2014social} found that frequent Facebook use is correlated with lower self-esteem, particularly when users compare themselves to others who appear more successful or happier on the platform. In a similar vein, Burke et al \cite{burke2020social} found that people who often compare themselves to others on Facebook tend to be highly active on the site and have larger friend networks. However, despite being exposed to more positive and socially engaging content, a substantial portion of these users experience negative feelings after encountering posts that make them feel worse about themselves. Other studies have also indicated that social comparison on social media has been linked to various negative outcomes such as depression \cite{samra2022social} and body dissatisfaction \cite{jiotsa2021social}.}
Pitfalls of social media-based online peer support have also been highlighted, such as personal distress due to others’ experiences~\cite{easton2017qualitative}, unhelpful interactions with others ~\cite{griffiths2015online}, social exclusion ~\cite{easton2017qualitative}, and feeling of vulnerability when talking to strangers online~\cite{breuer2015online}. Emerging evidence also documented other adolescent online safety issues including online sexual risks~\cite{hornor2020online, alsoubai2022friends, razi2023sliding}, exposure to explicit media content~\cite{chatterjee2023teen, park2023towards, rostad2019association}, and problematic media use with mental health issues ~\cite{mchugh2018social, park2023affordances, patchin2017digital}.


Given the disagreement within the literature, whether social media is beneficial or detrimental for adolescents remains an open question to explore that requires a nuanced picture of its potential benefits and detriments. The review and meta-analysis also showed that the association between social media use and teens' psychological well-being is still unclear as effects have been found to exist in both (positive and negative) directions~\cite{orben2020teenagers}. Yet, positive and negative effects can co-exist as effects can vary across different adolescents; some adolescents experience net positive effects, while others experience heightened risk ~\cite{odgers2022engaging}. For instance, through analyses of 2,155 real-time assessments of the adolescents’ unique susceptibility to the effects of social media, Beyens et al.~\cite{beyens2020effect} illustrated that the impact of passive (but not active) social media use on their well-being differed significantly from teen to teen, with 44.26\% of teens did not feel better or worse, 45.90\% felt better, and a small group felt worse (9.84\%). Therefore, efforts to holistically understand what adolescents are going through on social media are imperative to provide appropriate support systems for them and to inform design implications to provide a safer online environment, keeping in mind to maximize the benefits while mitigating potential risks and negative outcomes. This study takes this nuanced approach by analyzing teens' disclosures about their positive and negative experiences on various social media platforms.

\subsection{Analyzing Support Seeking Posts to Understand Teens' Lived Social Media Experiences} 

How youth seek and receive online social support has been an important area of study within the CSCW and HCI community. Previous research showed that adolescents go online to seek social support for sensitive topics such as mental health \cite{suzuki2004search}, romantic relationships \cite{kim2017romantic}, and sexual experiences \cite{razi2020let}. This is because adolescents are hesitant to receive professional help or talk to their parents about such sensitive topics ~\cite{forte2014teens, hartikainen2021safe} because of perceived stigma and embarrassment~\cite{radez2021children}. Most teens who face relationship difficulties do not seek help from people they know in person due to fear of judgment or concerns about confidentiality ~\cite{kim2017romantic}. Therefore, they turn to online social support to feel more comfortable interacting with people they do not know offline~\cite{ellison2016question}. Recently, researchers utilized digital trace data from social support platforms to understand adolescents' online social support-seeking behaviors~\cite{razi2020let, hartikainen2021safe, prescott2017peer}. 
For instance, by conducting a thematic content analysis of 622 initial posts shared by young people aged 11-25 years, Prescott et al.~\cite{prescott2017peer} explored how youth seek support on the online forum for emotional and mental health issues. They observed two distinctive ways of youths' online support-seeking: one was a direct request for advice, with a themed heading followed by a post with details about the specific advice they sought, while the other was finding other young people on the forum who shared similar experiences to themselves. Another main theme they found was that in many posts, youth shared their personal experiences and offered support for others ~\cite{prescott2017peer}. More recently, Razi et al. \cite{razi2020let} conducted a thematic analysis of over 4,000 teen posts on a peer support platform to understand how youth talk about their online sexual experiences. They found that most youths have utilized the platforms to seek support (85\%), while 15\% were more interested in connecting with others, and 5\% gave advice on different sexual matters. 
Although previous work has provided valuable insights into adolescents' support-seeking behaviors, most have focused on a particular type of negative online experiences (e.g. online sexual experiences, romantic relationship issues, digital stress) or risks experienced on a specific social media platform. 
Our work expands on previous studies by understanding adolescents' experiences on various social media platforms for which they seek online peer support and the differences in those experiences depending on different social media platforms. To do so, we analyzed posts that adolescents shared on an online social support platform for teens, where they explicitly mentioned the name of popular social media platforms.

\subsection{Leveraging an Affordances Perspective to Understand Teens' Differing Social Media Experiences} 


Meanwhile, adolescents use a variety of different social media platforms for different purposes including learning, entertainment, networking, and communicating with others~\cite{anderson2022teens, moreno2019applying}. 
For instance, teens use TikTok for entertainment, learning, and discovering trends, while, they use Snapchat for socializing, and go to Instagram for sharing pictures and checking what others are doing \cite{anderson2022teens}. Within the HCI literature, users' behaviors and experiences on social media platforms have been studied through the lens of “affordances," a relational structure between the social media platforms and the users~\cite{faraj2012materiality}. As a concept that captures the relationship between technology and users, affordances have been applied in social media research as a key concept to examine the perceived features of social media platforms that enable or constrain users' behavior on the platform~\cite{park2023affordances, ronzhyn2022defining}.  For instance, on social media platforms that value revealing real identities (e.g., Facebook), users tend to be more careful with how they present themselves, while on platforms that afford low identity affordances (e.g., Reddit), users disclose sensitive information without repercussions of personal identity~\cite{leavitt2015throwaway}. With their low identity affordances, (pseudo) anonymous platforms provided users with online spaces to seek social support regarding sensitive topics such as mental health, sexual encounters, and more. As such, affordances can aid us in understanding how adolescents interact on different social media platforms and in guiding them on the risks and benefits of each platform.

\edit{Prior research has revealed that user perceptions and interactions with social media platforms can vary significantly depending on the platforms' affordances. For example, affordances can influence the selection of social media platforms \cite{shane2018college}; Instagram is favored for its visual capabilities in sharing and enhancing photos and videos, Facebook for connecting with friends and family through personal interactions, and Twitter is implied to be appreciated for its real-time, concise communication ideal for disseminating information and participating in public dialogues. Another study comparing affordances between TikTok and YouTube platforms revealed that users prefer YouTube for its ability to control the sharing of information and manage who sees their information, activities, or profiles; on the other hand, TikTok was favored for helping users stay informed about information and the presence of others, allowing customization to suit personal preferences and interests, and making it easier to share content within the platform and to other platforms \cite{lee4557930comparative}. 
}
\edit{Furthermore, prior research suggests that individual differences affect how users interact with and perceive social media. One study \cite{devito2017platforms} found that people's traits, such as their experience with the platform and personality affect how they present themselves online. For instance, experienced users of a platform are likely to be more skilled at navigating the platform's features to manage their online presence more effectively. Personality also plays a role; individuals open to exploring diverse creative tools may express themselves more freely, while those higher in neuroticism may prioritize privacy controls and worry about negative feedback, affecting their online presentation \cite{devito2017platforms}.}


As such, previous literature provides an understanding of the concept of affordances and how it can be applied in social media research with general populations, yet the application of affordances in the adolescent online safety context remains under-studied. \edit{The motivation to explore social media affordances in the context of teens lies in understanding how the features of social media platforms shape teenagers' behaviors, experiences, and well-being. This exploration is important because teens may use social media differently than adults due to their unique developmental needs \cite{senekal2023social}, social dynamics \cite{jang2015generation}, and technological skills \cite{throuvala2019motivational}.} In this study, we examined teens' differing experiences across multiple social media platforms and interpreted some of our findings in light of the unique affordances of social media platforms. 
To do so, we explored the posts teens shared on an online peer-support platform with low identity affordances (i.e., registered users use pseudo-anonymous user names) to understand in what ways teens discuss social media platforms when they seek online support. By analyzing adolescents' posts on a pseudo-anonymous platform, we provide an understanding of adolescents' experiences on social media that they may not otherwise share with their parents or peers. In doing so, we investigate whether the way teens discuss social media platforms varies across different platforms. 




\section{Methods}

In this section, we first discuss platform selection and considerations made for data ethics. Then, we outline the methods for collecting data including a description of the data set and the data scoping procedure. Additionally, we present the data analysis approaches to answer our high-level research questions.

\subsection{\edit{Platform Selection and Considerations for Data Ethics}}
We licensed an existing dataset from an online peer support platform that aims to offer youth and young adults a safe environment to discuss topics such as mental health, sexuality, religion, and more. The name of this platform was anonymized to protect the teens' identities whose data we analyzed. On this platform, users can create posts, leave comments on posts, and choose whether to publish the posts anonymously or using their username. Hence, users have the ability to share their screen names and profile photos with other users when they intend, making their posts pseudo-anonymous; thus, safeguarding users' personal privacy. \edit{The platform is designed exclusively for sharing text-based posts, hence no images were collected in our dataset.} While the dataset did not include users' country of origin, most posts were in English. The licensed dataset consisted of about five million posts from 400 thousand users, including metadata (e.g., likes, hearts, and moods). The dates of the posts in the dataset ranged from 2011 to 2020. \edit{More details about the characteristics of our sample can be found in section \ref{4.1}.}

The Institutional Review Board (IRB) of our university determined that this study was exempt from the review for human subject research because personally identifiable information (e.g., usernames, contact information) was removed from the dataset prior to being shared with the research team. Nevertheless, we ensured that all authors completed their IRB Collaborative Institutional Training Initiative (CITI training) on human subjects research before accessing this dataset. We also took great care to remove any potentially personally identifiable information from the posts themselves (e.g., screen names for other platforms, hyperlinks) prior to sharing posts as exemplar quotations in the paper. 

\subsection{Data Scoping Process}

Our primary goal was to examine how teens (ages 13-17) discussed different social media platforms in their support-seeking posts. To accomplish this, we first ran a SQL query that searched for posts created by users whose ages were between 13 and 17 years old. Then, we narrowed our search criteria based on a keyword search of the ten most popular social media platforms among teens in the U.S., which included Instagram, Twitter, Facebook, Snapchat, YouTube, TikTok, WhatsApp, Reddit, Twitch, and Tumblr \cite{vogels2022teens}. Our search resulted in posts mentioning nine of the top ten platforms (Twitch did not appear in the search results). Also, to focus on more recent trends, we only included posts from the last two years of the dataset (2019 and 2020), which resulted in a total of \edit{ 2,465 posts. Some of these posts (\textit{n} = 205) were coded as irrelevant due to the alternative use of some of the keywords we used in our search, which can be found in Table \ref{tab:Keywords}. For example, the post ``\textit{well \textbf{ig} y'all don't understand}" appeared in our search results because of the word in bold.}
\edit{This post was deemed irrelevant because of the keyword “ig,” which in this context means “I guess”}. Lastly, after removing \edit{irrelevant (8\%)}, duplicate \edit{(7\%)}, and non-English posts \edit{(1\%)}, the final dataset contained  \edit{2,061} posts. We only focused on the original posts, excluding the comments on the posts, since we aimed to understand the support-seeking behavior of teens through their posts, rather than the support and advice they received as comments. 

\begin{table}[h]
    \caption{\edit{Scoping Search Terms}}
    \label{tab:Keywords}
    \begin{tabular}{|p{0.95\textwidth}|}
    \hline 
    \multicolumn{1}{|c|}{\textbf{Keywords}} \\
    \hline 
    \textbf{Social media platforms}: Instagram, Insta gram, Ig, Twitter, TW, YouTube, You Tube, YT, Snapchat, Snap Chat, SC, Reddit, Twitch, WhatsApp, Whats App, WA, Tumblr, Tiktok, Tik Tok, TT, Facebook, Face Book, FB \\
    \hline
    \end{tabular}

\end{table}

\subsection{Data Analysis Approach}
We first conducted a thematic analysis \cite{braun2012thematic} of teens' posts about their experiences on social media platforms (RQ1). \edit{To do this, we familiarized ourselves with the data by reading through the posts.} The first two authors then discussed the main topics presented in the posts to create initial codes. \edit{Subsequently, they divided the posts equally and coded using the same set of initial codes, while actively identifying new themes. Whenever the two authors identified potential new codes, other co-authors deliberated on whether to include them in the codebook.} If they reached an agreement, the first two authors went back and re-coded all previous posts to incorporate the new code. This process was iterative, involving constant communication and consensus among the researchers. \edit{Once all the data coding was completed, the first two authors worked with the last author to conceptually group the codes into cohesive themes that aligned with our RQ1. Finally, we reviewed our themes alongside our dataset to confirm that they actually captured important meanings within the coded data and named the themes. } \textbf{Table \ref{tab:codebook}} presents the themes and their corresponding codes along with illustrative quotations. Since we employed a double-coding approach for the posts, the total count of codes may exceed the overall number of posts. For example, teens sometimes mentioned more than one social media platform with varying experiences across the differing platforms. Therefore, the count of all codes can be totaled to be more than 100\% of our total post count.

After completing qualitative analysis (RQ1), we leveraged our codebook for statistical tests to identify patterns and trends in the data. While RQ1's qualitative analysis provided in-depth insights into teens' complex online experiences, a limitation lies in the challenge of integrating information across observations or assessing associations between observations~\cite{kirk1986reliability}. Conversely, the strength of the quantitative approach lies in its ability to compare observations and examine associations~\cite{castro2010methodology}. Our mixed-methods design, crucial for addressing RQ2 (relationship between platforms and teens' support-seeking behavior), combines qualitative and quantitative data~\cite{creswell2003advanced}. This approach facilitates accurate insights from confirmatory statistical analysis and deep explanatory descriptions derived from thematic analysis. For the qualitative examination of the relationship between social media platforms and teens' support-seeking behavior, we conducted $\chi^2$ tests, a between-group test for two or more variables~\cite{sharpe2015chi}. This method ensures that observed relationships between social media platforms and codes/themes from thematic analysis are not due to random chance, enhancing the rigor of thematic analysis findings. Prior research also supported such use of between-group tests in studies involving qualitative analysis of online content~\cite{unni2021shelter,basch2019content,harvey2019fear,lee2019sport}. To demonstrate significant differences between social media platforms, standardized residuals were used, calculated by "dividing the product of subtracting expected from observed values by the square root of the expected value"~\cite{sharpe2015chi}. Through these tests, we interpret nuanced differences between platforms based on youth disclosures of their experiences.

\edit{Prior to addressing our RQs, we first assessed for potential effects of the COVID-19 pandemic, which occurred in 2020. Therefore, we conducted a chi-squared ($\chi^2$) between the years 2019 and 2020 based on our main themes from RQ1. This analysis revealed no significant differences in the described themes in Table \ref{tab:codebook}, except for seeking information about social media. Standardized residuals showed that, during the pandemic, youth in our dataset were more likely to discuss seeking information related to social media platforms in 2020, compared with 2019. We highlight this difference in our results (section \ref{4.5}).} 
Next, we conducted a Chi-squared between-group analysis ($\chi^2$) to examine any significant differences between the online platforms based on the codes for each sub-theme (RQ2). We first categorized the posts based on the different social media platforms teens mentioned in their posts. Some posts (\textit{n} = 122) had multiple social media platforms mentioned, so we labeled the same post for each platform cited. \edit{For two of the social media platforms, Reddit and Tumblr, the code counts were notably small, with over 20\% of the codes having counts below five. This rendered our data inconsistent with the assumption of the $\chi^2$ test of having the expected values cells to be 5 or
more in at least 80\% of the cells~\cite{mchugh2013chi}. Therefore, we excluded these two platforms and conducted $\chi^2$ tests among the social media platforms that were more frequently discussed among the youth.}   
\begin{table*}
  \centering
  \footnotesize
\caption{Codebook for RQ1}
  \label{tab:codebook}
  \setlength{\tabcolsep}{7pt} 
\renewcommand{\arraystretch}{1.1} 
\resizebox{\columnwidth}{!}{ \begin{tabular}{| >{\raggedright}m{3cm} | >{\raggedright}m{3.5cm}  | m{8.5cm} |} 
\hline
\textbf{Themes/Subthemes} & \textbf{Codes/Subcodes} & \textbf{Illustrative Quotations} \\ 
 \hline


  \multicolumn{3}{|c|}{\textbf{The Negative Side of Social Media \edit{(58\%, \textit{n} = 1,202 posts)}}} \\ \hhline{|---|} 
 \multirow{5}{3cm}[10pt]{
 
\textbf{Negative Experiences from Social Media Use \newline \edit{(43\%, \textit{n} = 891)}}} & \textbf{Social drama} \edit{(22\%, \textit{n} = 461)} & \\

& - \textit{felt cheated \newline \edit{(11\%, \textit{n} = 230)}} & \textit{ 
`I have my boyfriend's account on TikTok and he made a comment on another girl's video. He also says goodnight to me and then stays online.” \newline }  \\ 
  & - \textit{relationships developed online \newline (6\%, \textit{n} = \edit{118})}  & \textit{ 
  `I accidentally was too nice to a man on WhatsApp (so he has my number) but I'm getting cold feet now bc he's an adult and I'm still considered a minor but I don't want to suddenly stop talking to him and he gets angry" \newline}  \\ 

  & - \textit{failed to get attention 
 \newline (5\%, \textit{n} = \edit{113})}  & \textit{ 
  `I feel so worthless when i take such good photos and post them on Instagram and get little likes while all my other friends get more and more and lots of followers while i lose about 1 a day and i always feel like giving up for good"}  \\ \hhline{|~--|}

  & \textbf{Cyberbullying}  \newline (4\%, \textit{n} = \edit{74})   & 
\textit{
`How tf [the fuck] you gonna follow my Instagram just to message me: " Nigga you weird", Seriously people have nothing better to do in their meaningless lives than to pick on and judge people and sit behind a phone screen and talk smack" } \\ \hhline{|~--|}

& \textbf{Body-shaming}   \newline (4\%, \textit{n} = \edit{74})     & 
\textit{
`The only thing girls are good for is judging me for showing my body on Instagram and ... calling me ugly this ugly ass ugly as fuck and hurting me" } \\ 

\hhline{|~--|}

& \textbf{Explicit content}   \newline (2\%, \textit{n} = \edit{32})    & 
\textit{
`Can someone tell me why random inappropriate content is always shown on my Snapchat discover feeds?" } \\ 
\hhline{|~--|}

& \textbf{Self-harm content} (1\%, \textit{n}=\edit{25})    & 
\textit{
`I just posted how to kill yourself on tumblr... I hope I dont get banned." } \\ 
\hhline{|~--|}

& \textbf{Interpersonal privacy violation}    \newline (1\%, \textit{n} = \edit{27})       & 
\textit{
`My mom always shares private stuff about me on Facebook. I really don't like that.
" } \\ 

\hhline{|~--|}

& \textbf{Other }(10\%, \textit{n} = \edit{198})       & 
\textit{
`I'm really scared to open Instagram rn but I need to”} \\ 
 \hhline{---}

 \multirow{5}{3cm}[30pt]{\newline 
 \newline
  \newline
  \newline
\textbf{Negative effects of using social media
\newline 
 (\edit{15\%, \textit{n} = 311)}}}

 & \textbf{Aroused anger}
\newline \edit{(5\%, \textit{n} = 111)}     & 
\textit{
`if i see anyone making a vid for Tiktok in public deadass im finna commit a murder" } \\ 

\hhline{|~--|}
 
 & \textbf{Harm to self-esteem } 
 \newline (\edit{3}\%, \textit{n} = 70)     & 
\textit{
`I completely abandoned Instagram for 3 months. ... Just now, when me and my cousin decided to be active again, scrolling though the posts kicked in all my insecurities again. And this time, it's worse than before" } \\

\hhline{|~--|}

& \textbf{Triggered mental health issues}  \newline (4\%, \textit{n} = \edit{83})    & 
\textit{
`My friend just posted an ugly face of me when I was a kid on Snapchat. Like he could've just posted other pictures. I look hideous. I'm so anxious about who is gonna see it. I started having panic attacks " } \\ 

\hhline{|~--|}

& \textbf{Disrupted time management  } \newline (2\%, \textit{n}= \edit{47})     & 
\textit{
`I watched youtube until I forgot that class is now ongoing" } \\ 

\hline


 
  \multicolumn{3}{|c|}{\textbf{Connecting on Social Media \edit{(45\%, \textit{n}  = 922 posts) }}}  \\ \hhline{|---|} 

 \multirow{5}{3cm}[35pt]{\newline 
 \newline
  \newline
  \newline
\textbf{ To boost online presence \newline 
 \edit{(26\%, \textit{n}  = 532)}}
} & \textbf{Invitation to follow each other} \newline \edit{(15\%, \textit{n} = 309)}   & 
\textit{
`who wanna be friends with me on tik tok my ushername is [User Name]" } \\ 
\hhline{|~--|}

& \textbf{Self-promotion} (11\%, \textit{n} = \edit{223})     & 
\textit{
`follow my instagram account [User Name] it's about depression, anxiety, low self esteem and many more!" } \\ 

\hhline{|---|}

 \multirow{5}{3cm}[30pt]{\newline 
 \newline
  \newline
  \newline
\textbf{To connect personally \newline 
 \edit{(19\%, \textit{n}  = 390)}}}
 
 & \textbf{Sought relationship} 
 \newline \edit{(10\%, \textit{n} = 216)}     & 
\textit{
`Anyone who won’t leave me on seen want to Snapchat and become friends? [User Name]." } \\ 
\hhline{|~--|}

& \textbf{Sought/offered emotional support}
\edit{(8\%, \textit{n} = 174)}     & 
\textit{
`my nan passed and i feeling nothing towards the death... I feel like I need to talk to someone. Is there anyone here to dm me?" } \\ 
\hhline{|---|}


 \multicolumn{3}{|c|}{\textbf{The Positive Side of Social Media \edit{(20\%, \textit{n}  = 409 posts)}}}  \\ \hhline{|---|} 

\textbf{Helped as coping mechanism} \edit{(11\%, \textit{n} = 234)} & \multicolumn{2}{|l|}{\textit{`When I feel low during studying I take some funny selfies in snapchat" }} \\ \cline{1-3} 

\textbf{Provided inspirational content}  \edit{(8\%, \textit{n} = 175)} & \multicolumn{2}{|l|}{\parbox{10.5cm}{\textit{`So I just watched at positive Memes and stuff on Instagram and THAT'S the selfcare that works for me" }}} \\ \cline{1-3}


 \multicolumn{3}{|c|}{\textbf{Seeking Information regarding Social Media \edit{(20\%, \textit{n}  = 421 posts)}}}  \\ \hhline{|---|} 

\textbf{Discussing social media content} (12\%, \textit{n} = \edit{247}) & \multicolumn{2}{|l|}{\textit{
`[YouTube link] is this actually true?.. like for real?” " }} \\ \cline{1-3}

\textbf{Social media features } \newline (5\%, \textit{n} = \edit{100}) & \multicolumn{2}{|l|}{\parbox{10.5cm}{\textit{`can you have more than one super bff on snapchat?" }}} \\ \cline{1-3}

\textbf{Technical challenges} \newline (4\%, \textit{n} = \edit{74}) & \multicolumn{2}{|l|}{\textit{
`anyone else Instagrams is messing up or just mine" }} \\ \cline{1-3}

\end{tabular}}
\end{table*}

\section{Results}
In our results, we first summarize the descriptive characteristics of our sample, followed by the statistical results (RQ2) across the social media platforms and the main themes that emerged from our qualitative analyses (RQ1). 



\subsection{Descriptive Characteristics of Our Sample}
\label{4.1}
Our study involved an analysis of \edit{2,061} posts generated by \edit{1,038} distinct users between the age range of 13 to 17 years. However, all the posts found were made by adolescents who fell within the age group of 15 to 17 years old. Notably, a significant proportion of the posts (70\%) were created by teens who were 17 years old, while about a third of posts came from those aged 16 \edit{(24\%)}, and only 1\% of posts came from those who were 15 years old at the time of posting. With respect to gender distribution, the majority of the posts were made by adolescents who identified as female \edit{(53\%)}, followed by those who did not specify their gender (33\%), and lastly, by those who identified as male \edit{(14\%)}. As shown in \textbf{Table \ref{tab:table2}}, the most mentioned social media platform in adolescents’ posts, Instagram was the highest with 35\% (\textit{n} = 720) of total posts. Snapchat ranked second with \edit{20\% (\textit{n} = 412)}. YouTube was the third most mentioned platform among teens with 14\% (\textit{n} = 282), followed by TikTok (\edit{10\%, \textit{n} = 207)}, Facebook \edit{(7\%, \textit{n} = 147)}, WhatsApp \edit{(\textit{n} = 127)} and Twitter \edit{(\textit{n} = 125)} with the same frequency level \edit{(6\%)}. Lastly, Tumblr \edit{(\textit{n} = 28)} and Reddit (\textit{n} = 13) were only mentioned in 1\% of the posts. Keeping in mind the relative percentages expressed by platform, there are some notable trends in \textbf{Table \ref{tab:table2}} to highlight. \edit{For instance, negative experiences were most frequently shared when teens referred to Instagram, Facebook, Twitter, Tumblr, and Reddit. Snapchat and Whatsapp had the highest relative proportion of posts related to using social media to connect with people, while YouTube had the largest proportion of posts related to information seeking. YouTube was the only platform where the number of positive posts exceeded the negative, possibly speaking to its overall popularity among youth.} \textbf{Tables \ref{tab:codebook} and \ref{tab:table2}} serve as the organizing structure for reporting our findings below. The $\chi^2$ results for RQ2 are presented at the beginning of each subsection and helped guide the selection of illustrative quotations presented in our qualitative results (RQ1). We use illustrative quotations to describe each theme and subtheme of our results. Each quotation is contextualized with the teen's age and gender information. 
\begin{table*}[]
 \footnotesize
 \caption{\edit{ How the ways of discussing different Social Media platforms changed across the platforms. * Percentages shown in each column are standardized to the total number of posts for that column to enable relative comparisons across social media platforms.}}
  \label{tab:table2}
\resizebox{\columnwidth}{!}{
\edit{
\begin{tabular}{lccccccccc}  
\hline
\textbf{Social Media Platform Type}   
 & \textit{Instagram} & \textit{Snapchat} & \textit{Youtube} & \textit{TikTok} & \textit{Facebook} & \textit{Whatsapp} & \textit{Twitter} & \textit{Tumblr} & \textit{Reddit} \\ \hline \hline
\textbf{Negative experience from social media use}
&\textbf{36\%*} &\textbf{28\%} &\textbf{21\%} &\textbf{20\%} &\textbf{42\%} &\textbf{29\%} &\textbf{29\%} &\textbf{36\%} &\textbf{38\%}  \\ 
Social drama   &18\%&14\%&12\%&13\%&21\%&19\%&18\%&14\%&9\% \\
Cyberbullying   &4\%&3\%&3\%&1\%&2\%&0\%&2\%&2\%&3\% \\
Body-shaming    &6\%&2\%&0.5\%&0\%&0\%&0\%&0\%&0\%&0\% \\
Explicit content    &1\%&2\%&0.5\%&1\%&1\%&1\%&1\%&2\%&0\% \\
Self-harm content     &1\%&1\%&0.5\%&0\%&0\%&1\%&0\%&5\%&3\% \\ 
Interpersonal privacy violation    &0\%&1\%&0.5\%&1\%&7\%&0\%&0\%&0\%&0\% \\
Other    &6\%&5\%&5\%&4\%&11\%&9\%&7\%&12\%&22\% \\
\hline
\textbf{
Negative effects of using social media
}   
&\textbf{11\%} &\textbf{6\%} &\textbf{10\%} &\textbf{28\%} &\textbf{8\%} &\textbf{6\%} &\textbf{10\%} &\textbf{19\%} &\textbf{6\%} \\ 
Aroused anger   &3\%&1\%&3\%&17\%&3\%&4\%&4\%&14\%&3\% \\
Harm to self-esteem   &4\%&2\%&2\%&2\%&2\%&1\%&1\%&0\%&0\% \\
Triggered mental health issues   &3\%&2\%&2\%&4\%&2\%&1\%&4\%&2\%&3\% \\
Disrupted time management   &1\%&1\%&3\%&5\%&1\%&0\%&1\%&2\%&0\% \\
\hline
\textbf{Social Media to boost online presence } &\textbf{23\%} &\textbf{20\%} &\textbf{15\%} &\textbf{16\% }&\textbf{7\%} &\textbf{10\%} &\textbf{23\%} &\textbf{12\%} &\textbf{19\%}  \\ 
Invitation to follow each other
&13\%&15\%&7\%&11\%&4\%&6\%&11\%&0\%&0\% \\
Self-promotion
&10\%&5\%&8\%&5\%&3\%&4\%&12\%&12\%&19\% \\
\hline
\textbf{Social Media to connect personally}  
&\textbf{13\%} &\textbf{29\%} &\textbf{3\%} &\textbf{4\%} &\textbf{8\%} &\textbf{33\%} &\textbf{6\%} &\textbf{0\%} &\textbf{0\%}  \\ 
Sought relationship  & 7\%&20\%&0\%&3\% &0\%&16\% &6\%&0\% & 0\% \\
Sought/offered emotional support   
&6\%&9\%&3\%&1\%&8\%&16\%&0\%&0\%&0\% \\
\hline
\textbf{Positive sides of social media} 
&\textbf{6\%} &\textbf{7\%} & \textbf{24\%}&\textbf{24\%} &\textbf{19\%} &\textbf{12\%} &\textbf{21\%} &\textbf{17\%} &\textbf{16\%} \\ 
Helped as coping mechanism
&1\%&5\%&11\%&14\%&13\%&3\%&17\%&12\%&13\% \\
Provided inspirational content   
&5\%&2\%&14\%&10\%&6\%&9\%&4\%&5\%&3\% \\
\hline
\textbf{Social Media to seek information}  
&\textbf{12\%} &\textbf{10\%} &\textbf{27\%} &\textbf{8\%} &\textbf{16\%} &\textbf{10\%} &\textbf{12\%} &\textbf{17\%} &\textbf{22\%} \\ 
Discussing social media content 
&5\%&4\%&23\%&5\%&6\%&7\%&3\%&12\%&16\% \\
Social media features
&3\%&4\%&2\%&3\%&6\%&2\%&7\%&0\%&0\% \\
Technical challenges
&4\%&2\%&2\%&0\%&4\%&1\%&1\%&5\%&6\% \\
\hline 
\hline
\textbf{Total number of posts: } &720 &412 &282 &207 &147 &127 &125 &28 &13 \\ 
\bottomrule
\end{tabular}}
}
\end{table*}
\begin{figure}[htp]
  \centering
  \includegraphics[width=0.7\textwidth]{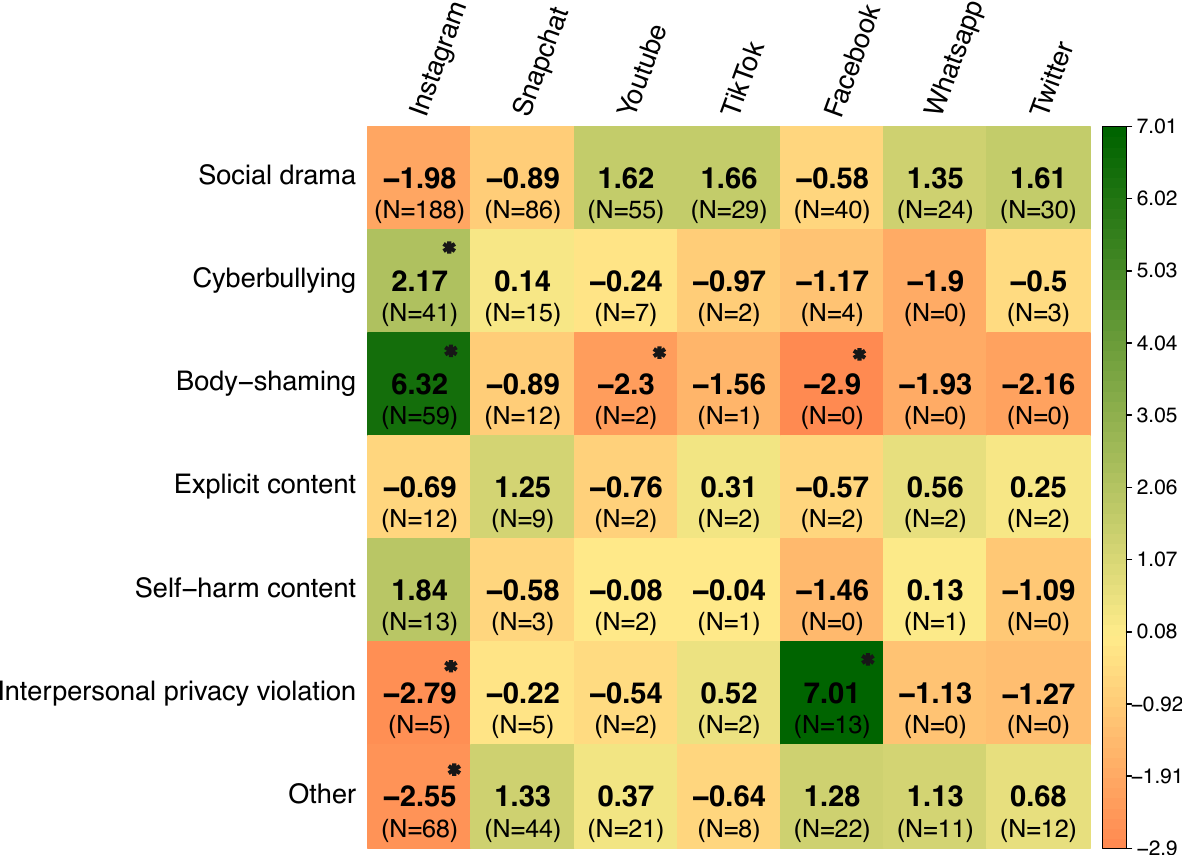}
   \caption{\edit{Results (standardized residuals) of the between-group analysis of social media platforms based on the teens' discussions about negative experiences when using social media ($N=891$). (*) indicates significant association. Note that green denotes a positive association, while red denotes a negative one.}}
   \label{NegExper}
 \end{figure}

\subsection{The Negative Side of Social Media}

We found that \edit{in 43\% of the posts (\textit{n} = 891), teens} mentioned different social media platforms \edit{included in our dataset} to share the negative experiences that they had online. The $\chi^2$ test indicated significant differences between these platforms based on relative expected proportions of the total number of posts \edit{$(\chi^{2} (df=36)=120.11, p < 0.001$)}. Below, we present how teens explained the adverse experiences they faced on different social media platforms.

\subsubsection{Negative Experiences from Social Media Use}
Teens \edit{(\textit{N} = 401)} shared many different kinds of unpleasant events that took place in their online lives. Among them, social drama was the most prominent. Around a quarter of the posts \edit{(22\%, \textit{n} = 461)} described different kinds of \textbf{social drama} that teens experienced from their social media feeds. Through these posts, adolescents often brought up the contentions, disputes, and strife they were going through with their social network, which often made them more stressed, sad, or upset. 
One of the most frequent types of social drama we saw is that teens often shared how their close social media connections made them \textit{feel cheated} \edit{(11\%, \textit{n} = 230)}. Teens most often felt cheated when their close social relationships (e.g., close friends, romantic partners) did not react or comment on their social media posts but rather engaged with someone else's posts. For instance, a teen shared her feeling of distress when her romantic partner commented on other girls' content on TikTok:  

\begin{quote}
\textit{``I feel physically sick. I'm almost certain my bf is using TikTok to follow and talk to other girls. He's just commented on this girl's post and liked it even though he shouldn't be saying those things because he goes out with me.''} - Female, 17 years old
\end{quote}


Teens also often felt deceived when they saw their friends or partners staying active online but not replying to their private messages on social media platforms. This indicates that adolescents assumed their social connections were active on social media by seeing their profile's active status, e.g., green dot on Instagram or Facebook Messenger. \edit{For example, a teen shared his disappointment with online friendships upon discovering that his friend had falsely claimed to be going to bed, only to find out that she was still active on Instagram.}

\begin{quote}
    \textit{``Fuck online friends. All they do is lie to me. This girl says she's heading to sleep on my Instagram but here I see her still online. Why do they all do this to me?"} - Male, 17 years old
\end{quote}

\edit{Teens often expressed a sense of betrayal from deception, particularly when close connections engaged more with others’ content or appeared active online but unresponsive to them, which triggered their need for validation and belonging. We saw that this need for validation, when unmet on platforms designed to foster social connections, exacerbated feelings of isolation and inadequacy. The digital visibility of these interactions, such as likes and comments, thus served as a double-edged sword, offering potential for affirmation, while also exposing vulnerabilities leading to feelings of neglect or rejection.}

Besides the above type of social drama, teens also posted about the \textit{relationships that they developed} \edit{(6\%, \textit{n} = 118)} on social media spaces that were potentially risky. Here, teens most often mentioned meeting strangers online who gave them a sense of importance by engaging with their social media posts through likes and comments and/or by messaging. We saw that most of these posts also depicted how these relationships, often romantic, instigated more emotional distress for them. \edit{For example, a teen shared her apprehensions after realizing that the individual she met on WhatsApp was considerably older than her and such an age difference could lead to potential complications in their relationship.} 


\begin{quote}
\textit{
``I talk to a guy on WhatsApp, I think he is way older than I am. I now realized it might create problems for me. This is making me worried sick. I dont want to things go messed up between him and I. Any suggestion on how to breakup with him but nicely?"} - Female, 16 years old

\end{quote}

\edit{The example above exemplifies the broader challenges some teens in our dataset encountered, where interactions with strangers initially seemed validating, especially in the absence of real-life connections. However, such interactions later became troubling, raising concerns about the safety and intentions of those involved. These types of posts highlight the critical need for awareness and caution in these interactions.}

In around 5\% \edit{(\textit{n} = 113)} of the posts, teens shared their sadness or distress regarding their social media follower counts. They created different digital content and shared them on their social media profiles, which \textit{failed to get enough attention} from their social media followers. They often mentioned that their shared content did not receive as many reactions or comments as other digital content creators usually would get, and they felt frustrated by this social comparison. \edit{For instance, in the following post, a teen compared the view counts of her own YouTube content with others' and shared how discontent she felt about it.} 

\begin{quote}
\textit{
``why is it that i work so hard to do well on YouTube and want to make ppl happy, but i can barely hit 100 views anymore. but ppl who post the most toxic videos get millions of views :(."} - Female, 17 years old


\end{quote}

\edit{The post above reflected a trend in the larger dataset, which illustrated how social media platforms tended to make teens feel like their social worth could be quantified. This emphasis on numeric metrics often led teens in our dataset to compare themselves to others unfavorably. We observed that such comparisons skewed self-perception and fueled a continuous search for validation through content creation, which often compromised authentic self-expression and overall well-being.}

Next, around 4\% \edit{(\textit{n} = 74)} of posts depicted teens' \textbf{cyberbullying} experiences on social media \edit{platforms included in our dataset}. As illustrated in \textbf{Figure \ref{NegExper}}, when comparing the proportions of teens' cyberbullying disclosures for each platform, we found that teens pinpointed Instagram significantly more frequently when they discussed cyberbullying experiences in their posts. Most of the teens' cyberbullying posts illustrated how teens were often harassed online regarding different content they shared in the form of images, videos, and/or texts on social media. Teens often mentioned that they received these hateful messages not just as comments on their social media feeds, but also as private messages, even from people they barely knew. 
For instance, a 17-year-old teen posted \edit{about a message received on Instagram containing offensive language and an aggressive tone:}


\begin{quote}

\textit{``So this bitch somehow messaged me on Instagram that im a pathetic little bitch and asks me but you wanna die you claim to be? Whoever the fuck you are Just fuck off"} - Female, 17 years old

\end{quote}

We also found that around 4\% of the posts \edit{(\textit{n} = 74)} were about different \textbf{body-shaming} comments teens received on the social media \edit{platfroms included in our dataset}. These teens mostly shared that their social connections made inappropriate or negative comments about their appearances, especially their body size or shape. For the body shaming posts, Instagram also had the highest relative proportion of teens disclosures about body shaming. This indicated that teens discussed their body shaming experiences on Instagram \edit{in their posts} more frequently than on other platforms \edit{included in our dataset}. 
The standardized residuals indicated that teens were statistically significantly more likely to discuss their body shaming experiences on Instagram and less likely to talk about these experiences on YouTube, Facebook, and Twitter (\autoref{NegExper}). 
\edit{The body shaming experiences shared by teens, though varied, were predominantly influenced by societal beauty standards. Many teens in our dataset noted that individuals who conform to these norms often receive praise and positive attention for their body images. However, when the teens themselves posted pictures where they felt confident but did not meet these standards, they encountered criticism and judgment.}


\edit{
\begin{quote}
\textit{`I guess i have to look good and really attractive before any girl talks to me. To them, i have to look really fit and good looking. Most of them just unfollow me on apps like IG and focus so much on my looks and body that they judge me and make me feel bad about myself saying my body is shit. Putting me down and making me insecure for showing it off on instagram. I fucking hate standards and shallow people 
"} - Male, 17 years old
\end{quote}
}



Apart from the above negative experiences, we found some posts that showed how teens perceived the \textbf{inappropriate content} they got exposed to when using the different social media platforms included in our dataset. About 2\% of the posts \edit{(\textit{n} = 32)} had complaints regarding different adult content teens found while exploring social media. We noticed that most of these posts are regarding other users who posted nude photos, implicating that some social media platforms allowed nudity. Here, majority of these posts illustrated teens' feelings of disbelief that some social media platforms with significantly large numbers of users lacked such content moderation. \edit{In one of such posts, a teenager questioned why a mirror selfie featuring someone showing their buttocks was considered acceptable and had not been removed from Instagram:} 

\begin{quote}
\textit{
``How the fuck was a mirror selfie of someone showing their ass on Instagram acceptable and not taken down? I'm so confused." } - Female, 16 years old
\end{quote}

Teens also sometimes encountered\textbf{ self-harm content} on social media \edit{platforms included in our dataset} and posted about it (1\%, \edit{\textit{n} = 25}). 
We observed that among all other negative experiences from social media use, teens were most distressed about this type of content. They mostly mentioned these contents showed someone's suicide or self-injury/cutting stories. We also noticed that teens complained about content that encouraged starvation or a fast diet to lose weight and explicitly identified such content as promoting self-harm. \edit{For example, a 17-year-old female expressed frustration with Instagram for not deleting a post uploaded by an internet celebrity who is well-known for extremely thin body, which she perceived as self-harm content.} 
\edit{
\begin{quote}\textit{
``Fuck you instagram for not taking down [celebrity name]’s posts while it’s clearly self harm and we all report it." } - Female, 17 years old 
\end{quote}}


We also found a few posts (1\%, \edit{\textit{n} = 27}) describing how using social media \textbf{violated teens' sense of digital privacy}. Most of these posts were about privacy breaches in which they felt their social media accounts were hacked or somehow their social connections viewed some of their posts that were set private. Teens also often complained about their close circles, e.g., families or friends, posting their photos or writing about their personal issues on social media without asking for their prior approval. Facebook had a higher relative proportion of posts discussing these experiences than other platforms \edit{in our dataset} as shown in Figure~\ref{NegExper}. This indicated that teens' posted more frequently about the violations that occurred to their interpersonal privacy on Facebook. When further examining these posts mentioning Facebook, teens mostly expressed that they felt their privacy was violated when their parents posted about them on Facebook without their consent (a.k.a "Sharenting" ~\cite{blum2020sharenting, steinberg2016sharenting}). For instance, the following quote describes a female teen's \edit{strong dissatisfaction with her mother sharing extensive details of her life on Facebook. She shared concerns particularly about her mother's constant updates related to herself, including her actions, current activities, and remarks, accompanied by ridicule, which made her feel a violation of her privacy.}


\begin{quote}
\textit{``It bothers me a lot when my mother posts about every single thing that happens in our life on Facebook. Especially about me (what I do, currently do and say, she ridicules me for it), I feel like it violates my privacy."} - Female, 17 years old
\end{quote}





To conclude, teens discussed different negative experiences they faced on social media. These posts mostly included complaints about social drama, cyberbullying, and body shaming. Sometimes, criticisms about social media platforms were shared without specific reasons. Next, we discuss the negative effects adolescents had when they used different social media platforms.

\begin{figure}[htbp]
  \centering
\includegraphics[width=0.8\textwidth]{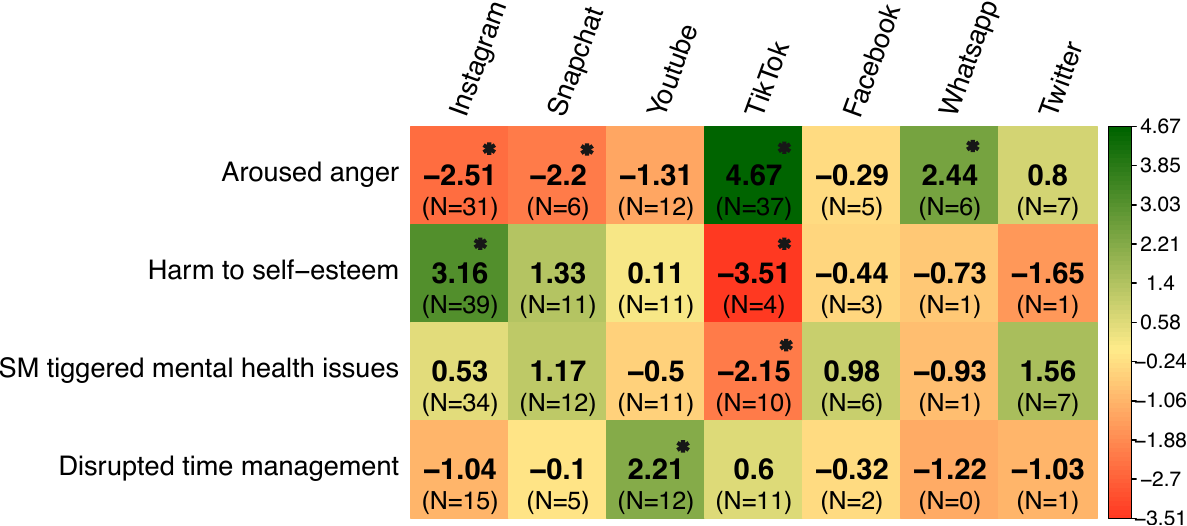}
   \caption{\edit{Results (standardized residuals) of the between-group analysis of social media platforms based on the teens' discussions about negative effects when using social media ($N=311$). (*) indicates significant association. Note that green denotes a positive association, while red denotes a negative one.}}
   \label{NegEffects}
 \end{figure}

\subsubsection{The Negative Effects of Using Social Media}

In addition to sharing about their negative experiences, \edit{183} teens also discussed the adverse effects they had from using social media (15\%, \textit{n} = 311). The $\chi^2$ test resulted in differences between the online platforms based on their discussions about the adverse effects of social media platforms \edit{($(\chi^{2} (df=18)=51.01, p < 0.001$)}. Within these posts, teens discussed different negative impacts, e.g., affecting self-esteem, mental well-being, and ability to manage their time effectively, that social media usage had on them. 

In around \edit{5\%} of the posts \edit{(\textit{n} = 111)}, teens shared their experiences of how various types of online content they encountered \textbf{aroused their feelings of anger}. Although the triggers for anger varied among teens, in most cases, their anger stemmed from disagreements with certain trends or challenges that they perceived as trivial or even potentially harmful. \edit{TikTok had the highest relative proportion of posts by teens expressing their anger followed by Instagram. In addition, the standardized residuals revealed significant positive associations between disclosures of aroused anger and TikTok and WhatsApp, as well as a significant negative association between these disclosures and Instagram and Snapchat (\autoref{NegEffects}). This indicated that teens had the highest probability to discuss their aroused anger experiences on TikTok and WhatsApp and had the least probability to post about this anger on Instagram and Snapchat}. Wihtin these posts, teens expressed frustration when confronted with content that did not align with their personal tastes, values, or interests. This often made them question the popularity of such content, creating a feeling of disconnection and subsequently leading to anger. For example, a feeling of frustration was expressed by a teen who complained about people creating TikTok videos that baited users to watch until the end for a big reveal that never came:

\begin{quote}
\textit{``Everytime I see those lame unfinished videos on TikTok, my blood boils!!"} – Female, 16 years old  
\end{quote}

Next, in about \edit{3\%, (\textit{n} = 70)} of the posts, teens mentioned how different online content they viewed \textbf{harmed their self-esteem}. By looking at \autoref{NegEffects} and comparing the relative proportion of teens' harm to self-esteem posts across the platforms, we found that Instagram had the highest relative proportion, reaching a positive statistical significance. This finding suggested that teens most frequently posted about harm to their self-esteem that occurred on Instagram, compared to other social media platforms included in our dataset. Through these posts, teens often mentioned how following individuals who they perceive as having a desirable lifestyle or attractive appearance led them to social comparison. Here, they mainly compared their own lives, achievements, and appearance to those showcased on social media, which sometimes resulted in feelings of inadequacy or a diminished sense of self-worth. Interestingly, we found some posts where adolescents \edit{shared the feelings of frustration and low self-esteem while blaming }their appearance for the low user engagement on their social media photos. \edit{For example, a teen shared about how insecure he felt about the appearance when others' photos on Instagram got more 'likes' than his own.}  

\begin{quote}
\textit{``You know I see all these guys and beautiful girls on Instagram and it gets me so uptight and mad cuz they are so good looking and get a lot of likes and I barely get anything cuz I'm so unattractive in person"}- Male, 16 years old 
\end{quote}


In about 4\% of the posts (\textit{n} = \edit{83}), teens indicated that certain content and features on some social media platforms were associated with \textbf{triggering their mental health issues}. In these posts, teens expressed how the public nature of some social media platforms amplifies personal insecurities, especially when content is shared without their consent. For instance, some teens mentioned how others posting unflattering photos of them was a potent trigger for stress and anxiety. In other cases, certain features of social media platforms contributed to retriggering pre-existing vulnerabilities and past traumas. \edit{For example, features like Snapchat's 'flashback memories,' which are designed to resurface past experiences, were reported to sometimes evoke unpleasant past events:}

\begin{quote}
\textit{``Those Snapchat flashbacks that pop up in your memories really got me fucked up tho"} – Female, 17 years old
\end{quote}

\edit{The example above, illustrates the broader issue of how teens in our dataset felt that social media algorithms are designed to maximize user engagement, often prioritizing content that elicits strong emotional reactions at the detriment of the users. Unfortunately, this approach, as observed in some of the teens' posts, led to the increased likelihood of teens encountering content that triggered anxiety, depression, or other mental health issues.}

We also found that in about 2\% of the posts (\textit{n} = \edit{47}), teens expressed the negative\textbf{ impact of social media on their time management.} Those teens highlighted their concerns about the distractions and addictive nature of social media platforms, which led to excessive time spent scrolling, watching, and interacting online. When comparing the relative proportions of teens' posts discussing their disrupted time management across platforms in our dataset, Instagram had the highest relative proportion of posts followed by YouTube. However, a statistically significant positive association was found between disrupted time management and YouTube as shown in \autoref{NegEffects}, indicating that teens had the highest probability to post about YouTube when they discussed their disrupted time management. The primary concern voiced by these adolescents centered around the challenge of successfully prioritizing tasks and maintaining a balanced approach to time management in the face of the continual appeal of social media. Within these posts, we observed that this is mainly due to the autoplay and recommendation features. For instance, one teen expressed how YouTube is distracting her from studying \edit{by continuously recommending interesting videos}:

\begin{quote}
\textit{``YouTube is the worst when I’m trying to concentrate on work, it keeps recommending some great videos"} – Female, 17 years old
\end{quote}

\edit{Many teens highlighted the cascading effects of poor time management and impulse control leading to addictive consumption of social media on their mental and physical health. The cycle of procrastination and constant distraction not only increased stress due to unfulfilled academic responsibilities but also led to a decline in academic performance. The habit of spending excessive time on screens significantly infringed upon their sleep quality, eroding both their mental and physical well-being.}

In sum, teens posted to vent about different unpleasant experiences they faced on social media, mostly because of the interaction with their social connections. They also often shared how the above negative encounters affected their mental health \edit{such as low self-esteem and recall of traumatic experiences. Within those posts, internet addiction and disruption in time management was another concerning trends.}  In the next section, we explore how teens leveraged the peer support platform to connect with others on social media.

 \begin{figure}[htbp]
  \centering
\includegraphics[width=0.8\textwidth]{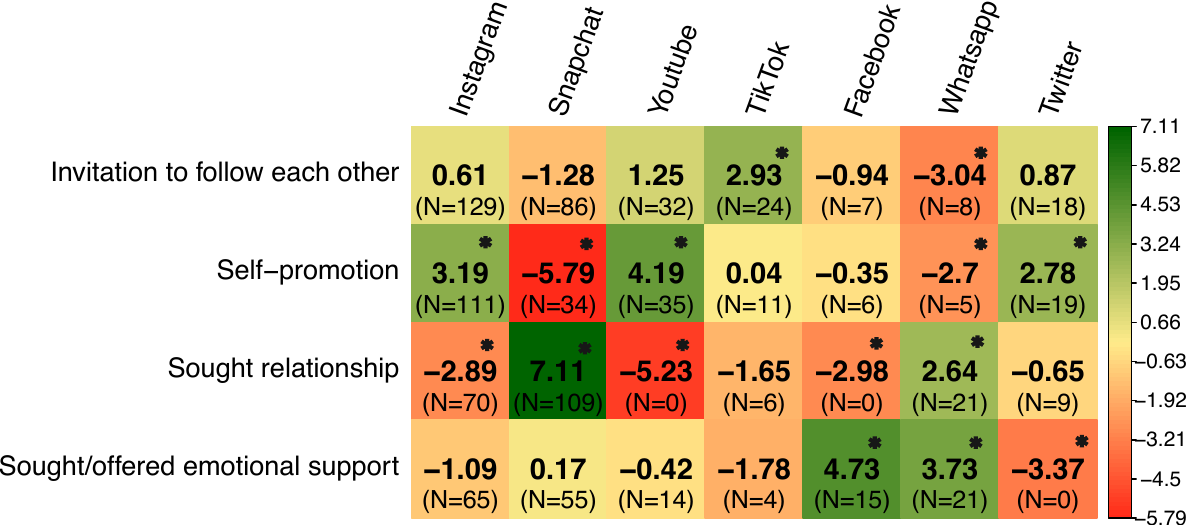}
   \caption{\edit{Results (standardized residuals) of the between-group analysis of social media platforms based on the teens' connecting on social media ($N=922$). 
   (*) indicates significant association. Note that green denotes a positive association, while red denotes a negative one.}}
   \label{connection}
 \end{figure}

\subsection{Connecting on Social Media}

Overall, we found that \edit{570} teens mentioned different social media platforms to connect with others \edit{in their posts (45\%, \textit{n} = 922)}. To this end, they invited others to connect on a variety of different social media platforms. Based on the $\chi^2$ test, there were significant differences between these online platforms based on the codes that emerged from teens' discussions about connecting with people theme \edit{($\chi^{2} (df=18)=153.53, p < 0.001$)}. Below, we present how teens invited others to increase their social connections.

\subsubsection{To Boost Online Presence }
We found that \edit{in 26\%} of posts \edit{(\textit{n} = 532), teens} cited social media \edit{platforms included in our dataset} as a way to boost their online presence. 
For example, in 15\% of the posts \edit{(\textit{n} = 309)}, teens shared their accounts and \textbf{invited others to follow them} in return for following them back. As illustrated in \autoref{connection}, Instagram had the highest relative proportion of posts related to teens' following invitations to others followed by Snapchat and YouTube. However, the standardized residuals showed a significant positive association between inviting others to follow with \edit{TikTok} and a significant negative association with WhatsApp. This suggested that when teens posted to invite others to follow them, they had the highest probability to mention \edit{TikTok} and the least probability to mention Whatsapp. \edit{For example, some teens encouraged others to follow their TikTok channels with the promise of following them back:}
\edit{
\begin{quote}
\textit{    ``If anyone has a TikTok feel free to put it in the comments and I will be sure to follow you back!"} – Unspecified, 17 years old
\end{quote}}


In 11\% of posts \edit{(\textit{n} = 223)}, teens tried to enhance their online presence through \textbf{self-promotion} of their social media accounts. Within these promotional posts, teens typically emphasized the content they provide in their accounts as a way to advertise themselves. Also, they often shared the links of their digital content and requested others to like, share, or comment. 
Based on the standardized residuals, 
teens more frequently posted about Instagram, YouTube, and Twitter to promote their profiles on these platforms (\autoref{connection}). 
Overall, we found that teens mentioned different social media platforms to enhance their online presence, mostly by 
exchanging their social media links and follow each other. 

\subsubsection{To Connect Personally}

Our analysis revealed that \edit{in 19\%} of posts \edit{(\textit{n} = 390)}, teens cited social media \edit{platforms included in our dataset} as a means to establish interpersonal connections with others through messaging. Unlike boosting the social media presence theme, these posts were mostly intended to invite others to personally talk or chat by sharing their user identifications on different social media platforms. 
For example, \edit{10\%} of the posts \edit{(\textit{n} = 216)} were about \textbf{seeking relationships} on social media that they could not establish offline. Within these posts, teens often expressed their desire to find the right partner, often displaying a sense of desperation, particularly in cases where their peers had already formed romantic relationships. As illustrated in \autoref{connection}, 
when teens posted about their seeking relationships on social media platforms, these posts had the highest probability to involve Snapchat and WhatsApp. For instance, a teen shared his Snapchat account with a specific desire to look for a romantic partner:

\begin{quote}
   \textit{``I need a girlfriend. When I see my friends with girlfriends, I feel so mad. Please if you are a girl, my Snapchat is [User Name]."}  - Male, 17 years old
\end{quote}

In addition to seeking relationships, teens also shared \edit{(8\%, \textit{n} = 174)} their social media accounts when they \textbf{wanted to exchange emotional support with others}. In these posts, teens often shared challenges, including depression, emotional detachment, and a desire for acceptance or personal connection with others. Also, we observed that teens often shared their social media accounts to offer others to listen
by sharing their own experiences and offering assistance to those who may be going through similar challenges. 
Instagram had the highest relative proportion of these posts followed by Snapchat, while based on the standardized residuals, 
teens' discussions about seeking or offering emotional support had the highest probability to involve Facebook or Whatsapp (\autoref{connection}). 
 \begin{figure}[!ht]
  \centering
\includegraphics[width=0.8\textwidth]{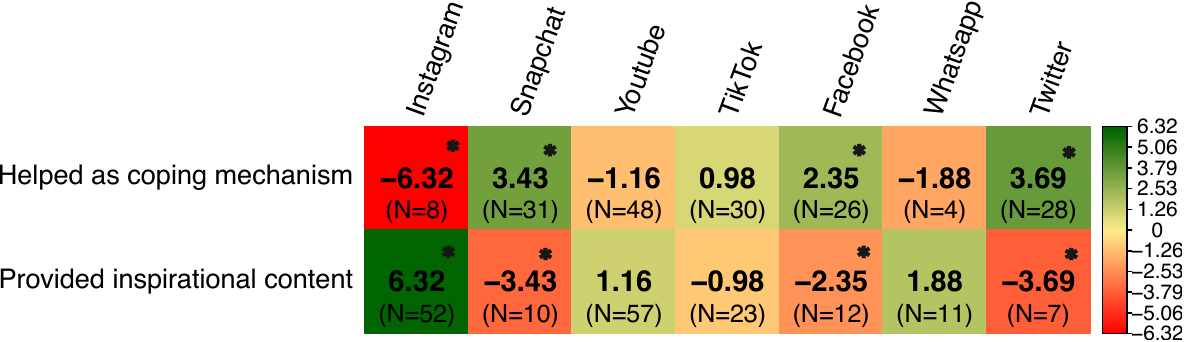}
   \caption{\edit{Results (standardized residuals) of the between-group analysis of social media platforms based on the teens' discussions about the positive sides of social media ($N=409$). (*) indicates significant association. Note that green denotes a positive association, while red denotes a negative one.}}
   \label{positive}
 \end{figure}

\subsection{The Positive Side of Social Media}
Meanwhile, \edit{151} teens talked in a portion of their posts \edit{(20\%, \textit{n} = 409)} about the positive aspects of different social media platforms. Statistically significant differences between the social media platforms based on the codes emerged from the positive sides of social media theme \edit{($(\chi^{2} (df=6)=65.72, p < 0.001$)}. About \edit{11\%} of the posts made by adolescents \edit{(\textit{n} = 234)} emphasized how they \textbf{utilize social media as a coping mechanism} during moments of low mood or emotional distress. \edit{YouTube had the highest relative proportion of posts describing social media as a coping mechanism, followed by Snapchat and then TikTok} as shown at \autoref{positive}. Yet, the standardized residuals 
showed that when teens posted about how they used social media as a coping mechanism, these posts had the highest probability to involve \edit{Snapchat, Twitter, and Facebook} and the least probability to involve \edit{Instagram}.   
Within these posts, some teens explained how engaging with others on social media platforms provided them with a sense of belonging and connectedness during hard times. For instance, some teens mentioned that they share funny selfies on Snapchat when they feel stressed during studying. Other teens found Snapstreak on Snapchat (i.e., tracks how many days in row users and their friends have exchanged Snaps and provides users extra points for their Snapchat score) rewarding enough to continue doing it as an important daily routine\edit{, as shown in the following post}. 

\begin{quote}
\textit{``This Snapstreak on Snapchat is life saving. It's the most important daily-routine of my life right now. Gives a purpose."} - Female, 16 years old 
\end{quote}

Others shared that consuming positive content on social media helped them relax when they felt stressed. For instance, some teens manage their Twitter accounts to push funny videos so that when they feel stressed, they can visit Twitter and ease their stress by watching those videos. \edit{For instance, in the following post, a teen shared that they filtered their Twitter content by setting preferences to display only humerous videos. When feeling sad, they turned to those curated feed of videos which served as a form of distraction or mood enhancement for them.}

\begin{quote}
  \textit{``I set my Twitter settings to have all funny videos and that helps me to stay away from all BS. When I am sad I just scroll through and that just works like magic."} – Unspecified, 17 years old
\end{quote}
\edit{As such, our analysis revealed that teens turned to social media for more than just posting content or engaging in Snapstreaks; they used it as a tool for emotional coping and seeking social support. These platforms provided spaces tailored to their emotional and social requirements. Moreover, it allowed teens to control their social interactions, empowering them in times of stress or isolation.}

In some cases \edit{(8\%, \textit{n} = 175)}, teens acknowledged social media as a valuable \textbf{source of inspiration}, motivating them through exposure to creative content, success stories, and inspirational quotes. YouTube had the highest relative proportion of posts describing social media as a source of inspiration. 
\edit{However, Instagram reached positive statistical significance} as demonstrated at \autoref{positive}. This suggested that teens more frequently included 
\edit{Instagram} in their discussions about how social media platforms provided inspiration and support. In their posts, teens most often expressed how this exposure helped them gain a fresh perspective when dealing with challenges and adversity in their offline or online lives. \edit{For example, some teens shared how they were motivated by the inspiring content they watched on Instagram}:
\edit{
\begin{quote}
\textit{``"Treat yourself like someone you loved." - some inspiring dude I saw on Instagram."} – Female, 16 years old
\end{quote}}



Overall, teens mentioned different social media platforms to appreciate the way they received support and inspiration from others and how social media, in general, helped them with their mental health. Next, we explore the ways in which teens sought information regarding different social media platforms on a peer support platform.

 \begin{figure}[htbp]
  \centering
\includegraphics[width=0.7\textwidth]{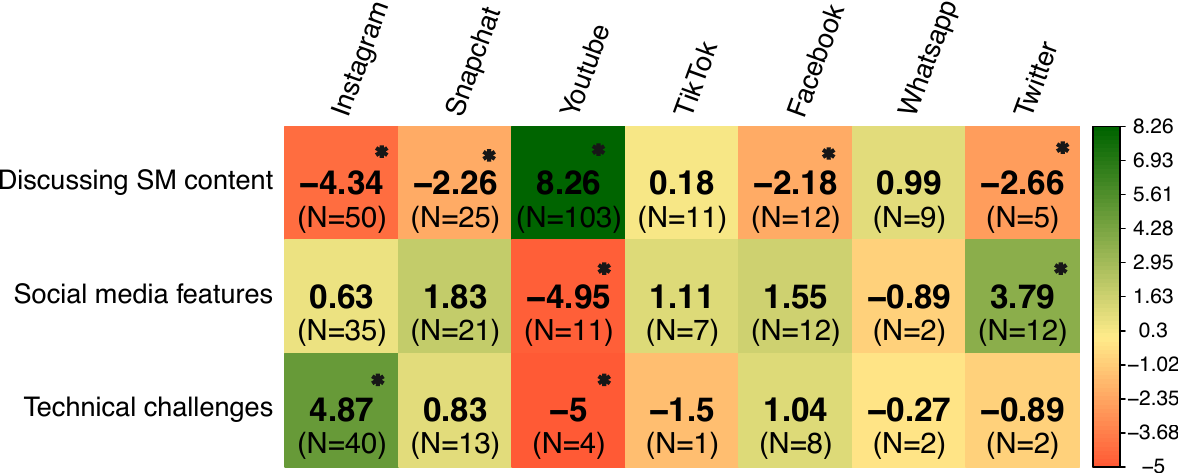}
   \caption{\edit{Results (standardized residuals) of the between-group analysis of social media platforms based on the teens' discussions about seeking information regarding social media ($N=421$). (*) indicates significant association. Note that green denotes a positive association, while red denotes a negative one.}}
  \label{seek}
 \end{figure}
\subsection{Seeking Information regarding Social Media }
\label{4.5}
As we examined posts, we found that \edit{235} teens aimed to seek information about various aspects of social media platforms in \edit{20\%} of their posts \edit{(\textit{n} = 421)}. \edit{Upon closer examination of our standardized residuals between 2019 and 2020, we saw that this trend has increased during the pandemic compared to 2019.} Overall, the $\chi^2$ test yielded significant differences between the platforms based on the emerged codes within seek information theme emerged from teens' discussions \edit{($(\chi^{2} (df=12)=92.86, p < 0.001$)}. For the most part, this was noticeable when teens \textbf{discussed social media content} (12\%, \edit{\textit{n} = 247}). As illustrated in~\autoref{seek}, YouTube had the highest relative proportion of teens' posts about discussing social media content across other platforms, showing also a positive statistical significance. This indicated that teens most frequently discussed content found on YouTube \edit{when seeking information about social media content.} Within this subset of posts, teens actively sought input from their peers by requesting recommendations for various social media content, e.g., YouTube videos, ranging from makeup tutorials to mental health support. \edit{Often times, they expressed a desire to confirm the authenticity and credibility of social media content. 
\begin{quote}
\textit{   ``[Youtube link] is this actually true?.. like for real?"} – Unspecified, 17 years old
\end{quote}}

Next, \edit{in 5\% of the posts (\textit{n} = 100), teens were seeking information on how to use different \textbf{social media features}.} 
When comparing the relative proportions of posts asking about features of social media platforms across all platforms, we found that Instagram had the highest relative proportions followed by Snapchat and then Instagram as shown at \autoref{seek}. \edit{However, based on the standardized residuals, there were significant associations between posts inquiring about social media features and Twitter. This indicated that teens' posts asking about social media features had the highest probability to involve Twitter.} We saw that many of these posts were centered around content creation, where teens posed questions regarding methods for creating fascinating and visually pleasing content. They primarily sought advice on employing filters, editing tools, and creative strategies to enhance their posts to appeal to their intended audience. In some other instances, teens sought assistance in understanding how to navigate \edit{privacy features (e.g., how to connect with others on specific social media platforms) to manage their online presence.}
\edit{
\begin{quote}
\textit{``im sorry, but how are you gaining an audience on an English-language twitter?"} – Unspecified, 16 years old
\end{quote}}


Lastly, a portion of the posts made by teens (4\%, \textit{n} = \edit{74}) centered around asking questions about different \textbf{technical challenges} they faced on social media. \autoref{seek} illustrated that Instagram had the highest relative proportion of posts discussing social media technical challenges in comparison with other platforms in our dataset, reaching also a positive statistical significance. This indicated that when teens seeking technical support, they most frequently referred to Instagram. In many cases, the technical challenges were about server problems or scheduled maintenance, resulting in temporary unavailability or restricted functionality on the platforms. In these instances, teens turned to the peer support platform to seek solutions and reassurances to navigate through the technical difficulties they encountered. \edit{For instance, a teen posted to inquire if anyone was experiencing issues with their Instagram, specifically mentioning a problem where the feed was not updating or changing as expected.}
\begin{quote}
\textit{``Somebody here has a problem with their Instagram? like the feed won't change."} – Female, 17 years old
\end{quote}

\edit{In summary, when teens mentioned various social media platforms in their posts, they predominantly posted to vent about different unpleasant experiences they faced on social media such as 
cyberbullying and privacy violation, while referring to Instagram, and Facebook. Furthermore, teens often shared how social media adversely impacted them in terms of their self-esteem, time management and arousing their anger 
particularly when mentioning Instagram, YouTube, and TikTok. In addition, teens often focused on connecting with others in their posts
for self-promotion and increasing their followers 
while referring to Instagram, YouTube, and TikTok. Also, teens expressed a need to communicate with others for emotional support and building online relationships when mentioning Snapchat, Whatsapp, and Facebook. 
On the other hand, teens also recognized and appreciated its positive aspects. For instance, they highlighted how social media served as a coping mechanism during difficult times and how it provided them with inspiration to strive for greater success, particularly citing 
Snapchat, Facebook, Instagram, and Twitter. Lastly, teens also sought information and guidance on how to effectively navigate the different social media platforms, mostly pointing to YouTube, Twitter, and Instagram.} In the next section, we discuss the implications of these findings.

\section{DISCUSSION}
In this section, we describe the implications of our findings in relation to prior work and provide design implications for implementing adolescent online safety features to help teens reduce their negative experiences on social media and maximize the benefits reaped. 

\subsection{A Balanced Narrative on Adolescent Social Media Use (RQ1)}



Our qualitative findings for RQ1 unpacked negative experiences that were indexed more frequently than the positive ones, as shown in \textbf{\autoref{tab:codebook}}. Yet, these seemingly overwhelming number of negative experiences might stem from teens' inclination to share during challenging times, especially since the context of the digital trace data analyzed was intended to garner peer-support. Therefore, we stress for the readers to avoid marking teens' social media usage using a false dichotomy of negative or positive. Instead, our focus should be on understanding how teens feel vulnerable to which online harms and for what underlying reasons so that we can provide appropriate support. This recommendation is well-aligned with prior research that emphasized studying the factors that make some teens at higher risk than others online~\cite{orben2020teenagers}. The impact of social media platforms on teens can vary from individual to individual, impacted by factors such as personality traits~\cite{rahardjo2020instagram}, family environment~\cite{vanwesenbeeck2018parents, akter2024familydesign}, and the overall digital environment~\cite{mccreanor2013youth}. 





Our research also highlights positive aspects of social media use among teens, such as coping with stress by sharing their content, connecting with others online, watching inspirational content to boost their mood and self-esteem, \edit{and seeking information during the pandemic}. Zooming out from the immediate findings, positive aspects of teens' social media usage were well-documented in the surveys and/or empirical studies (e.g., increased social connection, identity development, and positive emotions through social support)~\cite{anderson2022teens, haddock2022positive}. In addition, the meta-analyses confirmed that the association between social media use and teens' negative psychological well-being is still unclear, as effects have been found to exist in both (positive and negative) directions, and there has been little work done to rule out potential confounders~\cite{orben2020teenagers}. In a recent Pew Research survey, the most common way teens describe the impact of social media was neither positive nor negative~\cite{anderson2022teens}.  
Nevertheless, heightened attention to the negatives of teens' social media usage led U.S. Senators Blumenthal and Blackburn to propose the “Kids Online Safety Act (KOSA)” ~\cite{Blumenthal2022} to protect teens from online risks. The legislation requires social media platforms to proactively mitigate risks to minors. To do so, it requires independent audits and public scrutiny from experts to ensure that social media platforms are taking meaningful steps to address risks to kids~\cite{Blumenthal2022}. Some embrace the idea of KOSA given multiple claims that social media companies fail to protect teens~\cite{CNN}, while others fear that this legislation would incentivize social media sites to collect even more information about minors to prevent potential risks for them. 

Given the positive stories teens shared in our study and the trends found in existing research, we need to move away from over-emphasizing the harms that social media usage can bring to teens and be overly paternalistic about protecting them from potential harm. In the same Pew Research survey above, 22\% teens perceived that their parents are extremely worried about their social media life, while 39\% teens share that their online experiences are better than parents think~\cite{anderson2022teens}. Therefore, we once again highlight that social media usage is not necessarily positive or negative; rather, the two aspects co-exist without voiding one another, and net effects (positive or negative) can vary across different teens in various contexts. Therefore, further research is needed to understand differing teens' online experiences and how to amplify positive aspects at the same time tackle negative aspects to better support teens' digital well-being.

\subsection{Teens Pinpoint Concerns of Different Social Media Platforms Aligned with Their Affordances (RQ2)}

When teens discussed their negative experiences using social media, they often mentioned specific social media platforms. Specifically, our statistical analysis (RQ2) illustrated how teens mostly mentioned \textit{Instagram} whenever they discussed body-shaming and its harmful effect on their self-esteem, \textit{Facebook} when they talked about interpersonal privacy violations, and \textit{YouTube} when they discussed their struggles with time management (\autoref{NegExper} and \autoref{NegEffects}). To some extent, these statistical differences can be explained by the social media affordances, features, and social norms of these platforms that facilitate how users communicate, share information, and engage with others~\cite{zhao2013conceptualizing}. For instance, Instagram is a photo-focused platform, hence, a unique culture (e.g., users get more attention by posting physically appearing photos) was created on this platform that has strongly emphasized the significance of physical appearances \cite{tiggemann2018you}, which may affect teens as viewing some idealized pictures of others diminishes their self-esteem~\cite{casale2021multiple}.  \edit{Moreover, we saw posts in which teens described how 
following individuals whom they perceive as having a desirable lifestyle or attractive look on Instagram led them to social comparison and feel lower self-esteem, which has also been documented in prior research~\cite{jiang2020effects}}. Meanwhile, Facebook was launched over 20 years ago as one of the first platforms that allow users to use their real names~\cite{lee2021facebook}. Given this high identity affordances, family relationships are emphasized on this platform in a way that parents like to post about their teens~\cite{burke2013families}. These ``sharenting'' ~\cite{blum2020sharenting, steinberg2016sharenting} 
posts were found in our qualitative analysis as a source of frustration for these teens. On the other hand, YouTube offers a sheer volume of videos available for teens and employs mechanisms, such as autoplay, that make them feel less in control of the time they spend on the platform \cite{lukoff2021design},
which may result in their discussions about the long time they spent on this platform. While prior research has mainly focused on studying these affordances to understand the needs that would motivate people to use specific social media sites~\cite{karahanna2018needs}, our statistical analyses comparing the platforms based on teens' negative experiences discussions highlight the importance of adopting an affordances-based approach to address teens' negative experiences. Given the fast and evolving landscape of social media platforms, adopting this approach become of great importance for future researchers to provide a more nuanced understanding of how specific affordances contribute to the negative experiences of teens and/or other populations. Below, we abstract some of the key social media affordances that shaped teens' social media experiences, as described in our results:

\begin{itemize}

\item \textbf{Visibility of Status:} When status visibility was exposed, this often led to teens feeling cheated, overlooked, and excluded, especially when someone's words (e.g., ``\textit{I am going to bed.}'') did not align with their online activity (e.g., Green - ``Active Online'')

\item \textbf{Public Content Sharing:} Social media platforms that facilitated the public sharing of content, particularly image-based content (e.g., Instagram) amplified opportunities for cyberbullying, which was often perpetrated through private channels.

\item \textbf{Image-based Sharing:} Platforms that emphasized image-based sharing (Instagram) and selfies are more susceptible to negative experiences, such as body shaming.

\item \textbf{Attention-seeking or Status-signalling Affordances:} Attention or status-seeking affordances, such as follows, views, and likes, made teens feel undervalued or unpopular when their counts were lower than their peers or than their expectations.

\item \textbf{Self-Promotion versus Interpersonal Connection Affordances:} The affordances of some social media platforms (e.g., Instagram, YourTube, Twitter) were better suited for teen content creators who wanted to self-promote and create their own personal brands, while other platforms were better suited for forming interpersonal connections (e.g., Snapchat, WhatsApp). Differing privacy features afforded users to manage the audiences that they would like to interact with on each platform.  

\item  \textbf{Connecting Publicly with Like-minded Peers:} Social media platforms that afford social features (e.g., “Follow” and/or “Hashtag” on Twitter) facilitated users with building social connections by following each other and forming groups/communities of users with similar interests. 

\item  \textbf{Connecting Privately with Strangers:} When social media platforms allowed teens to connect with strangers privately (e.g., WhatsApp), this often led to the formation of regrettable relationships.

\item \textbf{Content Filters:} While most platforms had filters for inappropriate content, teens expected them to work better and expressed frustration when the content in their feeds (e.g., nudity) broke with this expectation. In contrast, when social media platforms afforded teens to create personalized content filters for what they wanted to consume versus avoid (e.g., sensitivity filter on Twitter), this contributed to more positive experiences for teens by lowering the chances of being exposed to unwanted content.


\item \textbf{Sharenting:} Some platforms (e.g., Facebook), likely due to real name affordances and the user audience, facilitated practices such as  ``sharenting,'' parental sharing of content related to their teens ~\cite{kumar2021oversharing,  moser2017parents}, which led to interpersonal privacy violations. 

\item \textbf{Ephemerality of Content:} Social media platforms that provided features where users could set how long their networks could view the content (e.g., Snapchat) had the potential of lowering teens' pressure to self-present and impression management, which has also been suggested by prior works ~\cite{piwek2016they, bayer2016sharing}.

\item \textbf{Curated Memories and Flashbacks:} Similar to the findings of prior work, social media platforms that curated past memories (e.g., Snapchat) had the risk of retraumatizing youth based on negative past experiences~\cite{scott2023trauma}, particularly for those who experience major changes in their lives~\cite{corvite2022social}. 

\item \textbf{Autoplay Content and Algorithmic Recommendations:} Social media platforms that used algorithms to recommend content and autoplay content (e.g., YouTube) had the potential of facilitating addictive patterns of consumption and time management struggles, which has been confirmed by prior research~\cite{lukoff2021design}.


\item \textbf{Other Engagement Features:} Features that promoted engagement (e.g., Snapchat streaks) served as a double-edged sword of promoting motivation and distraction from life's problems but also addictive behaviors.

\end{itemize}

In summary, we urge future research to adopt our affordances perspective when attempting to understand the differential patterns and consequences of social media use on youth. By taking this broader perspective, we can make a marked shift in our public discourse, research, design practices, as well as policies aimed at safeguarding youth from the detrimental effects of social media by moving towards a ``\textit{safety by design}''~\cite{terrede2022, agha2023strike, perrion2022} 
perspective that proactively considers how to shape the design-based affordances of social media in a way that is protective of youth without restricting or over-policing their online experience.

\subsection{Implications for Design}
Our study provides insight into the ways how teens discuss different social media platforms when interacting with other teens on online peer support tools, suggesting features and mechanisms for adolescents' online safety tools that would be needed to help them stay safe on social media. 

\paragraph{Design Affordances to Prioritize Youth Digital Well-being}

Given that some of the social media affordances that we identified above (Section 5.2) appeared to be problematic for teens, design-based recommendations can be drawn directly from our findings. For instance, as status visibility can negatively impact teens with feeling cheated or excluded, social media platforms can consider decreasing the visibility of status by providing options to manage the audience of their online status or turn online status off by default and choose whether they prefer to appear online.  
Also, as features such as views and likes can contribute to teens' feelings of low self-esteem and being left out, social media platforms can consider reducing some of the attention-seeking and/or status-signaling affordances. For instance, Facebook and Instagram are considering options to remove “Likes” for those under 18~\cite{facebook2019likes}. Other social media platforms with similar attention-seeking affordances can also consider the same design changes to reduce teens' pressure for social presence and attention-seeking on social media. Meanwhile, teens can become more susceptible to stranger danger on social medial platforms that afford teens to connect with strangers privately. Therefore, such platforms can create friction for connecting with strangers privately (e.g., WhatsApp, Instagram).  For instance, social media platforms can implement nudging systems to inform teens about the potential risks of viewing messages from strangers and to let them choose whether or not to view the message at all. Recently, Instagram implemented a policy in which adult users are not allowed to privately message teens under 18 who do not follow those adult users \cite{instagram2021restrictingdms}; proactive interventions such as this can help social media platforms to actively moderate online stranger danger.

Additionally, as autoplay, recommendation algorithms, and other engagement features (e.g., Snapchat Streaks) can contribute to addictive patterns of online content consumption and time management struggles. In fact, addictive media use and consumption have become one of the salient social issues to the extent that a bill that requires social media companies to take measures to mitigate the risks of internet addiction, the “Social Media Addiction Reduction Technology Act (SMART Act),” being proposed by the U.S. Senate ~\cite{congress2019SMART}. The bill was to prohibit social media companies to exploit users with addictive features such as YouTube autoplay and SnapStreaks which make it difficult to leave a social media platform~\cite{guardian2019ban}. With these new regulations, social media platforms are starting to introduce new safety features (e.g., features that allow users to customize screen time limits for each day) that can help control their media use~\cite{tiktok2023newfeature}. Along with the SMART Act bill and new safety features, we call for social media platforms to provide teens with more safety options including turning off or customizing autoplay and other engagement features. More ideally, we suggest design recommendations to support teens' healthy social media usage by promoting self-regulation and autonomy, which we explicate in detail in the following section.   

\paragraph{Toward Intentional and Meaningful Social Media Use}
Our statistical analyses showed that teens are more likely to refer to YouTube when they discuss disrupted time management. That is, video-sharing social media platforms tend to keep teens engaged to a level they feel is unhealthy, hence, teens need more agency to set healthy boundaries (e.g., screen time) when on social media. In fact, video-sharing platforms are known to intentionally employ a variety of design mechanisms (e.g., autoplay and recommendation) to maximize users' engagement with the platforms; even big-tech industry insiders warn that many of these mechanisms are designed to exploit psychological vulnerabilities and negatively impact the users ~\cite{lukoff2021design}. These practices that exploit human psychology to substantially impede freedom of choice led the US Senate to propose the SMART Act~\cite{congress2019SMART}. 
One way to manage screen time is through the approaches of intentional and planned use of social media platforms (especially multimedia-based ones). Previous research documented that users can make active choices to control their screentime on YouTube when they have a specific intention in mind; the more specific these intents (e.g., to watch a certain video on a certain topic), the stronger agency to control their screen time~\cite{lukoff2021design}. 
The intentional and planned use of media is known to be especially effective in promoting self-regulation from an early childhood~\cite{hiniker2017plan}. With the initial prompt from parents, teens are capable of learning intentionality and making goal-directed choices as planned, which is the mediating factor in developing self-regulation~\cite{hiniker2017plan}. 
Therefore, social media platforms, particularly multimedia-based platforms such as YouTube or TikTok could consider adding a screentime planning feature where teens can indicate their intention and set screentime themselves before using the platform, and reward them for achieving pre-planned screentime. This will allow teens to navigate healthy boundaries that work best for them. Also, it will ease tension between parents and teens, especially those who struggle with negotiating conflicting boundaries. 

\paragraph{Designing Context-Aware and Intelligent Safety Features with Teens}
Our results showed that teens are still being exposed to risky content (i.e., explicit and/or self-harm content) even though there are filtering and/or reporting features in place on social media. This suggests that the safety features do not reflect contextualized teens' risk experience, largely because while risk is highly subjective, safety features are designed based on the risk perceptions of the third person~\cite{park2024personally}.  Hence, we need to look again at those safety features through the lens of teens' perspective. One way to do so is to actively include teens in the design of safety features of social media through participatory design that considers teens as the primary stakeholders of their own online experiences. For instance, 
\edit{the social media platforms} where teens are more likely to encounter self-harm or explicit content can have co-design sessions with them to understand how they use their existing reporting tools and how to improve the usability of such tools. \edit{One potential resolution may involve improving the visibility of these reporting tools.} They can also design new user interfaces with teens such as real-time nudges to alert those who try to post self-harm content to think twice before they post them. \edit{Yet, the alerts approach introduces an important and ongoing debate about balancing the risks associated with self-harm content online against the critical need for a supportive space where youth can express and manage their struggles safely~\cite{lavis2020online}. Therefore, it is essential to develop nuanced policies and designs that focus on how to safeguard young users while facilitating their access to help and empathy. One way to understand the interplay between the risks of exposure to harmful content and their need for supportive spaces could be conducting co-design sessions with teens who could provide valuable insights on what types of alerts, content filters, or support provisions would be most effective for them.}
 In addition, by giving agency to the design process of their online safety solutions, teens can reflect on their own online habits and learn how to self-regulate those habits in ways that promote resilience, autonomy, and digital well-being~\cite{chatlani2023teen, ali2024case, agha2023co}.
Therefore, we encourage social media developers and designers to actively work with teens to include their unique perspectives when designing safety features for promoting their online safety.   

\paragraph{Towards Teens-Led Efforts for New Data Collection to Train Online Risk Detection Algorithms} In this paper, we identified salient negative experiences and effects that teens discussed \edit{when referring to} \textit{different} social media platforms. \edit{Our results indicate that algorithmic approaches to identify online risks on social media for youth could be more teen-centered and effective if they take into account the difference between the platforms and the types of risks teens discuss encountering the most on those platforms.} In contrast, prior works on machine learning (ML) algorithms for detecting social media risks for youth have mainly centered on training models using \textit{available} datasets ~\cite{razi2021human, kim2021human, alsoubai2024systemization} without considering teens' shared experiences on these platforms. Therefore, instead of focusing our efforts on collecting benchmark datasets from various social media platforms, we recommend prioritizing addressing platform-specific challenges that teens discussed when they disclosed their negative experiences on these platforms. Collecting data from platforms that align with teens' risk experiences would enhance the relevance and realism of the training data. For instance, there have been efforts to publish benchmark datasets from several social media platforms such as Facebook, Twitter, Instagram, and Reddit to train detection models for body shaming~\cite{gasparini2022benchmark}. However, in our study, we found that adolescents were most likely to discuss experiencing this risk on Instagram. Therefore, to make the data collection efforts more focused and teens-centered, we suggest using data from Instagram to train automated algorithms to detect teens' body shaming instances. \edit{Drawing from the Razi et al. case study~\cite{razi2022instagram}, where researchers successfully gathered and analyzed similar sensitive data from social media platforms, it is evident that with appropriate methodologies and ethical considerations, collecting targeted data from Instagram to detect teens' body shaming instances is indeed feasible.} By doing so, the actual risks faced by teenagers in their online lives would be acknowledged, which would result in allowing the risk detection models to accurately identify these risks on specific platforms (e.g., \cite{park2023towards, razi2023sliding, ali2023getting}). As such, the models would learn patterns, contextual cues, and platform-specific dynamics that would be more representative of the risks faced by them. 
\edit{At the same time, we acknowledge the transferability of ML algorithms across different platforms is difficult and should be done with caution~\cite{koh2021wilds, blackwell2018online, correa2015many}. Our results show from an empirical perspective that the teens' risk experiences across various platforms differ, hence, could inform that algorithmic approaches should be tailored to different platforms based on their affordances and the types of risks teens discuss encountering the most on those platforms.} 

\paragraph{Collaborative Online Safety Tools for Adolescents}
Our results reconfirm that adolescents experience online risks on social media and shed light on how such adverse experiences affect their mental health to the point that they seek out support due to these experiences. At the same time, our results confirmed that teens felt their privacy was violated when their parents became involved in their lives via social media. One potential approach to address the problem of privacy violations that occur via sharenting may be to provide collaborative features for family members, where teens could proactively review their parents' posts \textit{before} they are shared via the platform, and vice-versa. 
 Through such interactive feedback, parents and teens can better understand each others' privacy boundaries. 
 A large body of prior research on online safety domain suggested moving away from restrictive and paternalistic approaches \cite{wisniewski_parental_2017, ghosh_safety_2018}, as they cause more fear and paranoia among families \cite{pain_paranoid_2006}, and moving toward adopting more collaborative \cite{akter_evaluating_2023, akter_CO-oPS_2022, kropczynski2021examining} and teen-centric approaches \cite{ghosh_circle_2020, akter-from-2022, hashish_involving_2014, akter_it_2023}, where adolescents can work alongside their parents to make their online safety decisions together \cite{badillo2020towards} while having some level of personal privacy \cite{cranor_parents_2014}. \edit{However, these privacy-preserving collaborative approaches may still be open to conflicts between the two parties. For example, parents could post excluding the child from viewing, or teens could intentionally block parents' benign posts. A suggested collaborative mechanism to address potential conflicts is the Circle of Trust introduced by Ghosh et al. in \cite{ghosh_circle_2020}. This mechanism enabled parents to only see inappropriate message contents exchanged by teens within their social network and allowed them to negotiate with teens about excluding specific individuals from their trusted circle.  We urge future researchers to explore privacy-preserving yet effective ways to mitigate the potential concerns of such parent-teen collaborative mechanisms.}  
Thinking beyond the parent-teen relationship, however, our research also highlights the importance of \textit{peer relationships}, as teens in our study sought advice and support regarding their social media experiences from strangers on an online peer support platform. This is consistent with prior work that shows youth may prefer seeking support regarding sensitive topics from peers, or even strangers \cite{rayland2023social}, rather than their parents. As such, it might be time to change the status quo of relying on parents as the mediators of adolescent online safety to considering ways in which we can engage peer support as a protective mechanism that teens already leverage. Thus, we urge design practitioners to consider such collaborative and peer-based approaches when designing online safety features, so that teens can reach out to their peers for help when navigating different social media risks.

\subsection{Limitations and Future Work}
There are several limitations of our work that inform future research directions. First, we were not the first to uncover many of the themes that we identified from our analyses, as issues like cyberbullying, body-shaming, and privacy violations have been well-documented by prior research \cite{kwan2020cyberbullying, reddy2022you, de2020contextualizing}. However, our research is one of the first to take an affordances perspective to better understand youths' differentiated social media experiences across different platforms. As such, our work serves to both confirm many of the findings from previous work and build upon and add nuance to those findings, as well as shift the conversation towards a more practical design-based approach to proactively address some of the problems youth encounter when using different social media platforms. Second, our sample was skewed toward older adolescents (ages 15-17) who identified as males, females, or who did not specify their gender. Furthermore, these teens were active users on an online platform dedicated to peer and mental health support. Therefore, our results may not be generalizable to the whole teen population of different sexual orientations, ages, and those not actively seeking peer support pseudo-anonymously online. Future work needs to triangulate our results across broader and more diverse demographics of youth to ensure that our results hold. Additionally, we caution readers to not assume that a higher percentage of negative posts provides evidence that the drawbacks of social media use outweigh the benefits. Prior research has confirmed negative bias in online reviews \cite{qahri2022negativity}. Therefore, we assume that youth were more likely to share their negative experiences than the good ones. 




Another key limitation of our work is that we did not analyze the support and advice teens received from others based on comments made on their posts. Such analyses would be helpful in understanding potential recommendations for coping the negative experiences, amplifying positive ones, and generally the helpfulness of such advice. Since 10\% of posts were removed before coding because they were duplicates, this suggests that these teens may have received fewer responses than they sought. Finally, our analyses were constrained by the time period available for analyses (2019 and 2020) as the social media landscape has since changed. Yet, a key strength of our work is in demonstrating for the purpose of future research the importance of examining the lived experience of teens across different social media platforms and doing so based on digital trace data in places where they are actively seeking support based on these experiences.

   \section{Conclusion}
We explored how teens discuss social media platforms in their support-seeking posts on an online mental health platform. Four primary themes were identified in the discussion of social media platforms, including connecting on social media, negative sides of social media, positive sides of social media, and seeking support/information on social media. Our analysis also showed the ways teens discuss these themes vary across different platforms, along with the unique social media affordances of each platform. We also provide important design and policy implications to make experiences on social media more inspiring and less detrimental for teens. The key takeaway from our study is that the benefits and drawbacks of teens' social media usage can co-exist and net effects (positive or negative) can vary across different teens in various contexts. Therefore, we need to move away from fear-based approaches where we try to overly protect teens from potential harm. Rather, the focus should be on how to empower teens with positives, at the same time, tackle negatives to better support teens’ digital well-being. 

\begin{acks}
This research was supported by the U.S. National Science Foundation under grants IIP-2329976, IIS-2333207, and the William T. Grant Foundation grant 187941. Any opinions, findings, conclusions, or recommendations expressed in this material are those of the authors and do not necessarily reflect the views of our sponsors.
\end{acks}
\bibliographystyle{ACM-Reference-Format}
\bibliography{main}










\end{document}